\documentclass[onecolumn]{aastex631}

\usepackage{natbib}
\usepackage{url}
\usepackage{amsmath}
\usepackage{CJKutf8}
\usepackage{hyperref}
\usepackage{graphicx}
\usepackage{booktabs}
\usepackage{rotating}
\usepackage{amssymb} 
\usepackage{bm}      
\usepackage{lmodern} 

\shortauthors{Deng et al.}
\begin{document}
\begin{CJK}{UTF8}{gbsn}

\title{The Multi-wavelength Extinction Law and its Variation in the Coalsack Molecular Cloud Based on the Gaia, APASS, SMSS, 2MASS, GLIMPSE, and WISE Surveys}

\correspondingauthor{Shu Wang, Biwei Jiang}
\email{shuwang@nao.cas.cn , bjiang@bnu.edu.cn}

\author[0009-0000-1055-4355]{Juan Deng (邓娟)}
\affiliation{School of Physics and Astronomy, Beijing Normal University, Beijing 100875, People$’$s Republic of China}
\affiliation{Institute for Frontiers in Astronomy and Astrophysics,
	Beijing Normal University,  Beijing 102206, People$’$s Republic of China}
\affiliation{CAS Key Laboratory of Optical Astronomy, National Astronomical Observatories,
	Chinese Academy of Sciences, Beijing 100101, People$’$s Republic of China}
\affiliation{Department of Astronomy, China West Normal University, Nanchong 637000, People$’$s Republic of China}

\author[0000-0003-4489-9794]{Shu Wang (王舒)}
\affiliation{CAS Key Laboratory of Optical Astronomy, National Astronomical Observatories,
	Chinese Academy of Sciences, Beijing 100101, People$’$s Republic of China}

\author[0000-0003-3168-2617]{Biwei Jiang (姜碧沩)}
\affiliation{School of Physics and Astronomy, Beijing Normal University, Beijing 100875, People$’$s Republic of China}
\affiliation{Institute for Frontiers in Astronomy and Astrophysics,
	Beijing Normal University, Beijing 102206, People$’$s Republic of China}

\author[0000-0003-2645-6869]{He Zhao (赵赫)}
\affiliation{Purple Mountain Observatory and Key Laboratory of Radio Astronomy, Chinese Academy of Sciences, 10 Yuanhua Road,  Nanjing 210033, People$’$s Republic of China}

\begin{abstract}
Accurate interpretation of observations relies on the interstellar dust extinction law, which also serves as a powerful diagnostic for probing dust properties.
In this study, we investigate the multi-wavelength extinction law of the quiescent, starless molecular cloud Coalsack and explore its potential variation across different interstellar environments: the surrounding region, the nearby high Galactic latitude region, the inner dense region, and the inner diffuse region.
Using a sample of 368,524 dwarf stars selected from Gaia DR3 as tracers, we establish the effective temperature $T_\mathrm{eff}$-intrinsic color relations to derive the intrinsic color indices and optical$-$mid-infrared (MIR) color excess (CE) for 20 bands. 
Linear fits to the CE–CE diagrams provide color excess ratios (CERs), which are subsequently converted into relative extinction.
The resulting extinction curves for different environments exhibit steep slopes in the near-infrared (NIR) and flat profiles in the MIR. In the optical-NIR range, the Coalsack extinction law is consistent with $R_\mathrm{V}$ = 3.1 of \cite{2019wangshu} and \cite{2020H&D}, while in the MIR it follows $R_\mathrm{V}$ = 5.5 of \cite{2001wd5.5} similar to the results of active star-forming clouds. At an angular resolution of ${1.3'}$, our extinction map reveals fine cloud structures. No correlation is found between $R_\mathrm{V}$ and $E_\mathrm{B,V}$ for $E_\mathrm{B,V} >$ 0.3 mag, implying a uniform optical extinction law in the Coalsack cloud. The derived average $R_\mathrm{V}$ value is 3.24 ± 0.32.

\end{abstract}


\section{Introduction} 
The interstellar-dust extinction law, or extinction curve, characterizes the wavelength-dependent absorption and scattering of light by interstellar dust grains. Accurate determination of this law is crucial for uncovering the intrinsic properties of reddened objects, interpreting observations correctly, and revealing the composition, size distribution, and structural properties of dust within the interstellar medium (ISM) \citep{2003Draine}. The absolute extinction value at a given wavelength $\lambda$ is defined as $A_\lambda$, which is difficult to calculate directly. Instead, two common indicators of extinction are used: the color excess ratio $E(\lambda - V)/E(B - V)$ and the relative extinction $A_\lambda/A_\mathrm{V}$, where $E(B-V)$ is the color excess \citep{1989ApJCardelli}. In the Milky Way, ultraviolet to optical extinction law can be characterized by a total-to-selective extinction ratio $R_\mathrm{V}$ ($R_\mathrm{V}  \equiv  A_\mathrm{V}/E(B-V)$), which depends on the interstellar environments and varies significantly along different sight lines \citep{1989ApJCardelli,1999Fitzpatrick}. 
The average value in Galactic diffuse regions is $R_\mathrm{V} \approx 3.1$ \citep{2003Draine,2019wangshu}. $R_\mathrm{V}$ can range from $\sim$ 2 in low-density regions \citep{1999Fitzpatrick,2017WangShu} to $\sim$ 6 in dense molecular clouds \citep{1990Mathis,1999Fitzpatrick,2019Fitzpatrick}. The near-infrared (NIR, 0.9–3 $\mu$m) extinction law generally follows a power-law of $A_\lambda \propto \lambda^{-\alpha}$, with $\alpha$ varying between 1.6 and 2.4 across different sight lines \citep{2014ApJWang,2015Schultheis,2018SSMatsunaga}. Recently, \cite{2024Wangshu_JWST} established a nonparametric $\alpha$-dependent extinction curve for the wavelength range of 0.6–5.3 $\mu$m, providing extinction coefficients for the JWST NIRCam bandpasses. However, the mid-infrared (MIR, 3-8 $\mu$m) extinction law remains less well-understood. 
Early studies, such as \cite{1985ApJ...288..618R}, suggested that the NIR power-law might extend into the MIR bands, with extinction laws outside dense molecular clouds appearing uniform. \cite{1990Martin&Whittet} identified a universal extinction curve extending from the NIR to at least 5 $\mu$m for both diffuse and dense clouds. Later studies found that in dense interstellar environments, the MIR extinction law tends to be flat, consistent with the extinction law of $R_\mathrm{V} = 5.5$
(\citealp{2001wd5.5}, hereafter \hyperref[WD01_label]{WD01})\label{WD01_label}.

Dense ISM regions with high extinction are  crucial for understanding extinction law. Molecular clouds, cradles of star and planet formation, are typically dense regions with significant reddening and important targets for studying extinction law. Numerous investigations into extinction across various molecular clouds, including Rho Ophiuchi, a molecular cloud complex near the cluster IC 5146, Taurus,  Serpens, Trifid, Pipe, Orion, Perseus, California, and the Eagle Nebula’s Pillars of Creation show that the optical extinction law often agrees with $R_\mathrm{V} = 3.1$, while the infrared extinction law with $R_\mathrm{V} = 5.5$ \citep{1978ApJ...226..829H,1994ApJ...429..694L,2001Whittet,2005Knez,2009Chapman,2021Chu,2023Cao,2023Li,2023Lil,2024Li}. However, most of these studies focus on the active star-forming clouds, neglecting quiescent ones. While the quiescent molecular clouds show no clear signs of star formation, they still possess dusty, complex environments. Whether the extinction laws derived from active star-forming clouds can be applied to quiescent clouds remains uncertain. 

To address this issue, we select the Coalsack, a prominent representative of quiescent molecular cloud \citep{2008Nyman}, as our target. Despite its importance, the extinction properties of the Coalsack cloud remain poorly characterized. Previous studies, including \cite{1989southernCoalsack}, \cite{2005Dobashi}, and \cite{2022Guomap}, measured its extinction values for specific band, and \cite{2013Wangshu_coalsack} probed only the infrared extinction law in selected sub-regions. However, a comprehensive analysis of the extinction law across the entire Coalsack could is still lacking. With the advent of extensive, high-precision photometric and spectroscopic data from surveys like Gaia, investigating multi-wavelength extinction laws and its variation across the entire cloud region is now both feasible and essential. In this work, we use Gaia DR3 \citep{2023Gaia3} to select dwarf stars as tracers and gather photometric data across 20 bands from multiple surveys to establish effective temperature $\mathit{T_\mathrm{eff}}$–intrinsic color relations for deriving intrinsic colors. The color excess (CE) is calculated for each star, and the color excess ratio (CER) is obtained by linear fitting of two CEs. The CER is then converted into relative extinction  ($A_\lambda/A_\mathrm{V}$ or $A_\lambda/A_\mathrm{K_S}$). Finally, we present the optical-MIR reddening and extinction laws for the Coalsack molecular cloud and explore its potential variation across different interstellar environments, including the surrounding region, the nearby high Galactic latitude region, the inner dense region, and the inner diffuse region.

This paper is structured as follows. Section \ref{sec:data_samples} introduces the data and samples selection. Section \ref{sec:Method} describes the methods for deriving multi-wavelength CERs and relative extinction. Section \ref{sec:Results and Discussion} presents the results, and explores the optical-MIR extinction law and its potential variation. Section \ref{sec:Summary} concludes the main results.

\section{Data and Sample}\label{sec:data_samples}

\subsection{Coalsack Molecular Cloud}\label{Coalsack}
The Coalsack is one of the most prominent starless clouds in the southern sky, easily visible to the naked eye, yet it remains among the least studied nearby molecular clouds \citep{2008Nyman,2011Beuther}. Located in the Galactic plane around $\mathit{l}$ = 303°, $\mathit{b}$ = 0°, it spans an on-sky size of about 6° \citep{2008Nyman}, with \cite{2024MagneticCoalsack} suggesting it may extend up to 10°. Internal variations of $A_\mathrm{V}$ are significant, particularly in small condensations and globules, and can exceed 20 mag in some regions \citep{1973Tapia,1977Bok}. \cite{1988Gregorio} derived its average $A_\mathrm{V} \approx$ 5 mag. 
The Coalsack cloud is likely a young cloud complex in the early stage of evolution, characterized by a complex filamentary structure and numerous dark cores, with a total mass of about 3500 $\mathrm{M_{\odot}}$ \citep{2008Nyman}. 
Its complexity and vast size encompass diverse environments, from opaque dense cores to translucent and diffuse regions, making it an ideal target for studying extinction law variation across different interstellar environments \citep{2013Wangshu_coalsack}.
Distance estimates for the Coalsack molecular cloud show slight variations: $\mathit{d}$ = 180 ± 26 pc \citep{1989southernCoalsack}, $\mathit{d}$ = 188 ± 4.1 pc \citep{1989A&ACoalsack}, $\mathit{d} \approx$ 150 pc \citep{2008Nyman,2023Dharmawardena}, $\mathit{d}$ = 200 pc \citep{2024MagneticCoalsack}，but consistently confirm it as a nearby object.

\subsection{Data}\label{alldata}

Dwarf stars are used to trace the clouds' extinction because they are numerous and their parameters are well determined. Based on the stellar parameters from the Gaia catalog, we establish a dwarf sample. By cross-matching with the APASS, SMSS, 2MASS, GLIMPSE, and WISE catalogs with a radius of ${\,1''}$, the photometric data across 20 bands are collected, spanning from optical to MIR wavelengths.

Gaia DR3, the third data release from the Gaia mission, provides precise photometry in three broad bands: $\mathit{G_\mathrm{BP}}$,  $\mathit{G_\mathrm{RP}}$, and $\mathit{G}$ \citep{2023Gaia3}. The $\mathit{G_\mathrm{BP}}$, $\mathit{G_\mathrm{RP}}$, and $\mathit{G}$ bands span the entire optical range from 330–680, 630–1050, and 330- 1050 nm, respectively \citep{2018gaia}. 
It also includes the General Stellar Parameterizer from Photometry (GSP-Phot) results, detailing essential stellar parameters such as $\mathit{T_\mathrm{eff}}$, surface gravity (log $\mathit{g}$), and metallicity ([M/H]) for 471 million sources \citep{2023Andrae}. Notably, the $\mathit{G}$ band is excluded from this study due to its significant variation in the extinction coefficient with stellar $\mathit{T_\mathrm{eff}}$ and extinction \citep{2018Danielski,2019wangshu,2023Lil}.

The SkyMapper Southern Survey (SMSS) provides photometry in six filters: $\mathit{u}$, $\mathit{v}$, $\mathit{g}$, $\mathit{r}$, $\mathit{i}$, and $\mathit{z}$, with central wavelengths at approximately 350, 384, 510, 617, 779, and 916 nm, respectively \citep{2011Bessell}. The sky coverage extends from the South Celestial Pole to $\delta$ = +16°, with some observations reaching $\delta$ $\approx$ +28° \citep{2024smssdr4}.
We use photometric data from the latest release DR4, which includes over 700 million objects with significantly enhanced astrometry and photometry in comparison with previous version \citep{2024smssdr4}.

The American Association of Variable Star Observers Photometric All-Sky Survey (APASS) provides photometry in five filters (Johnson $\mathit{B}$, $\mathit{V}$, and Sloan ${g'}$, ${r'}$, ${i'}$) for stars \citep{2014apassbv}. We use data from the latest release DR9 \citep{2016passdr9}. 
Since the quality of the $\mathit{g}$, $\mathit{r}$, and $\mathit{i}$ band data from SMSS is better than the ${g'}$, ${r'}$, and ${i'}$ band data from Sloan, only the $\mathit{B}$ and $\mathit{V}$ data are taken from APASS.

The Two Micron All-Sky Survey (2MASS) provides comprehensive NIR photometry, covering more than 300 million objects \citep{2003Cohen}. We utilize the $\mathit{J}$ (1.24 $\mu$m), $\mathit{H}$ (1.66 $\mu$m), and $\mathit{K}_\mathrm{S}$ (2.16 $\mu$m) photometric data from its point-source catalog \citep{20062mass}.

The Galactic Legacy Infrared Mid-Plane Survey Extraordinaire (GLIMPSE) is a Spitzer Space Telescope Legacy Science Program, surveying the inner Galactic plane in infrared with the IRAC camera at wavelengths centered on 3.6, 4.5, 5.8, and 8.0 $\mu$m \citep{2003Glimpse,2009Churchwell}. We use four IRAC bands: [3.6], [4.5], [5.8], and [8.0]. 

The Wide-field Infrared Survey Explorer (WISE) provides full-sky photometry in four MIR bands: $\mathit{W}$1, $\mathit{W}$2, $\mathit{W}$3, and $\mathit{W}$4, with central wavelengths of 3.35, 4.60, 11.56, and 22.09 $\mu$m, respectively \citep{2010wise}. We extract photometric data from the AllWISE catalog. Since few sources in our sample have reliable W4 magnitudes (with uncertainties $<$10\%), we only use data from $\mathit{W}$1, $\mathit{W}$2, and $\mathit{W}$3 bands.

\subsection{The Samples} \label{sec:Samples}

\begin{figure}
\centering
\includegraphics[width=0.7\textwidth]{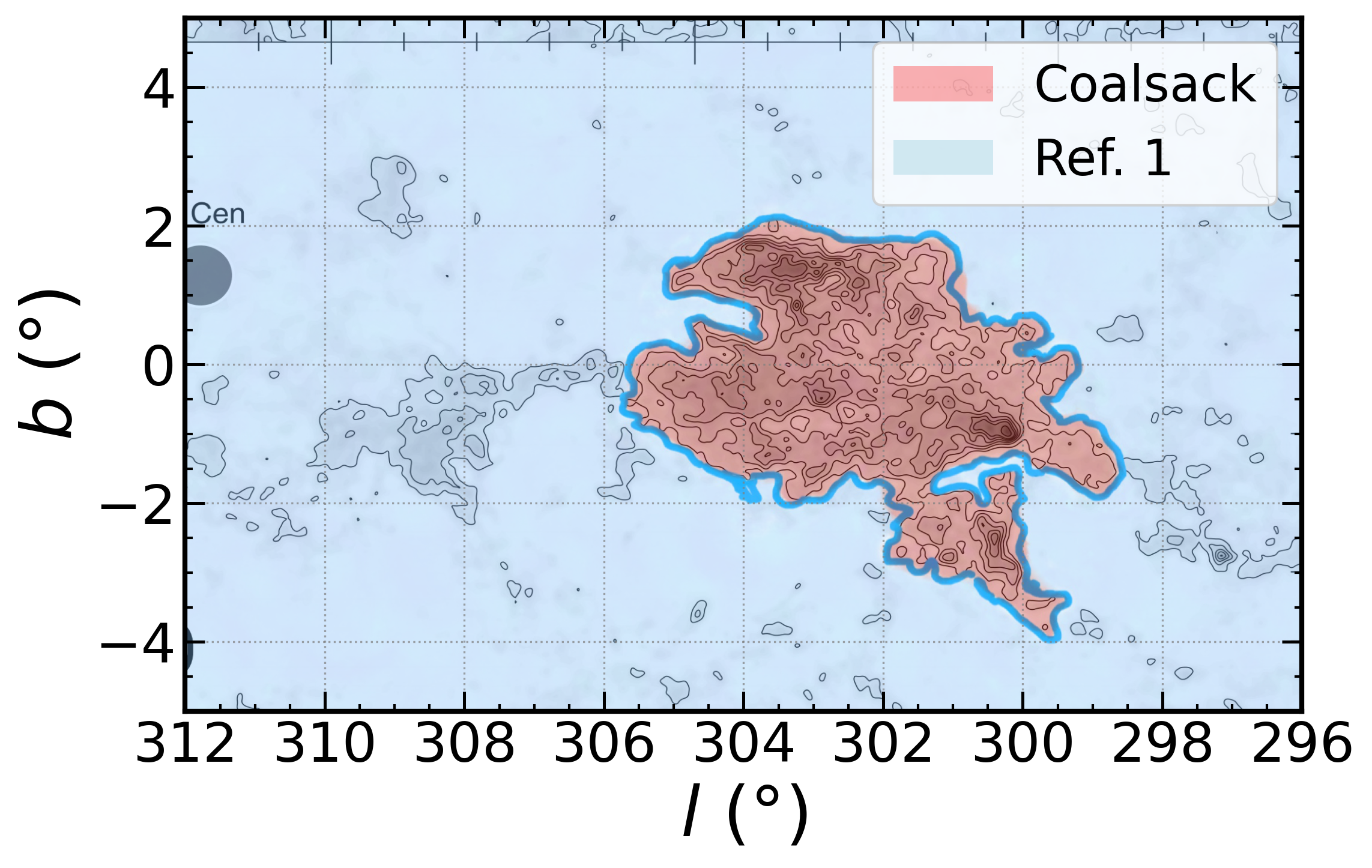}
\caption{The spatial map of the samples covers the Galactic coordinates \(296^\circ \leqslant l \leqslant 312^\circ\) and \(-5^\circ \leqslant  b  \leqslant 5^\circ\), based on the $A_\mathrm{V}$ contour map from \cite{2005Dobashi}. 
The Coalsack cloud region is shaded in red, while reference region 1 (Ref. 1) is marked in blue.
\label{fig:Coalsack}}
\end{figure}

The Coalsack region is selected within \(299^\circ \leqslant l \leqslant 306^\circ\) and \(-4^\circ \leqslant  b  \leqslant 2^\circ\) and highlighted in red in \autoref{fig:Coalsack}. The selection is based on the $A_\mathrm{V}$ map from \cite{2005Dobashi}  that provides sharp contours with 0.5 mag intervals for $A_\mathrm{V} < 3.0$ mag and 1.0 mag for $A_\mathrm{V} > 3.0$ mag. 
For comparison, two reference regions with different extinction depth are selected.
One is the region of $296^\circ \leqslant l \leqslant 312^\circ$ and $-5^\circ \leqslant b \leqslant 5^\circ$ excluding the cloud, to represent the surrounding cloud-free  environment, labeled as Ref. 1 (hereafter), shown in blue in \autoref{fig:Coalsack}. 
In addition, a more diffuse region adjacent to the cloud is selected with \(296^\circ \leqslant l \leqslant 305^\circ\) and \(5^\circ \leqslant  b  \leqslant 10^\circ\) labeled as Ref. 2 (hereafter). 

To reduce the influence of background clouds, the dwarfs are cut within 1 kpc to trace the Coalsack extinction. They are selected by stellar parameters from Gaia's GSP-Phot. The GSP-Phot provides stellar parameters from $\mathrm{BP}$/$\mathrm{RP}$ spectra using low-resolution blue and red prism photometers \citep{2023Andrae,2023Creevey}, validated by \cite{2023Fouesneau}. The median absolute error in $\mathit{T_\mathrm{eff}}$ and $\log g$ is 119 K and 0.2 dex, respectively.
For this work, the uncertainties are restricted by $\sigma_\mathrm{T_\mathrm{eff}}$ $< 180$ K, $\sigma_\mathrm{log \mathit{g}}$ $< 0.2$ dex, and $\sigma_\mathrm{[M/H]} < 0.5$ dex. The dwarfs are identified candidates in the $\mathit{T_\mathrm{eff}}$-$\log g$ diagram with $\log g \geq 4$, $4000 \leqslant \mathit{T_\mathrm{eff}} \leqslant 8000$ K, and $-1.0 \leqslant$ [M/H] $\leqslant 0.5$ dex. The final Coalsack, Ref. 1 and Ref. 2 samples contain 4,757, 117,585 and 32,964 stars, respectively, as listed in \autoref{tab:samples}. 

The selected data are cross-matched with the photometric catalogs detailed in Section \ref{alldata} and the following selection criteria for each catalog to ensure data quality: 
\begin{enumerate}
    \item For Gaia, photometric error $< 0.05$ mag,  $\mathit{G_\mathrm{BP}}$ and $\mathit{G_\mathrm{RP}}$ $< 18.0$ mag.
    \item For APASS, photometric error $< 0.05$ mag in $\mathit{B}$ and $\mathit{V}$ bands. 
    \item For SMSS, $\mathit{x}\_flags < 4$, $\mathit{x\_nimaflags < 5}$, and $\mathit{x\_ngood} > 0$, where $\mathit{x}$ refers to $\mathit{u}$, $\mathit{v}$, $\mathit{g}$, $\mathit{r}$, $\mathit{i}$, and $\mathit{z}$ bands.
    \item For 2MASS, photometric error $< 0.05$ mag and magnitude ranging from 6.0 - 14.0 mag in $\mathit{J}$, $\mathit{H}$, and $\mathit{K}_\mathrm{S}$ bands.
    \item For GLIMPSE, photometric error $< 0.1$ mag in [3.6] and [4.5] bands, $< 0.2$ mag in [5.8] and [8.0] bands.
    \item For WISE, photometric error $< 0.1$ mag in $\mathit{W}$1, $\mathit{W}$2, $\mathit{W}$3 bands.
\end{enumerate}

\section{Method} \label{sec:Method}
\subsection{Intrinsic Color Index} \label{subsec:Intrinsic Color Index}
The blue-edge method is used to derive the intrinsic color index ($C^0_{\lambda_1,\lambda_2}$) of stars \citep{2001Ducati}. This method assumes that the bluest stars have little to no extinction in a large stellar population, serving as zero-reddening proxies. Their observed colors ($C^\mathrm{obs}_{\lambda_1,\lambda_2}$) represent $C^0_{\lambda_1,\lambda_2}$ for a set of stars with specific parameters. \cite{2014ApJWang} derived the $\mathit{T_\mathrm{eff}}$–$C^0_{\lambda_1,\lambda_2}$ relationship for K-type giants (3500 K $\leqslant \mathit{T_\mathrm{eff}} \leqslant $ 4800 K). Subsequently, $C^0_{\lambda_1,\lambda_2}$ values for various types of stars have been determined across multiple wavelengths \citep{2016XueMengyao,2017WangShu,2018SunMingxu,2023wangshu}. Additionally, \cite{2024zhoahe2} used machine learning method for $C^0_{\lambda_1,\lambda_2}$ calculation based on the blue-edge method. The blue-edge method relies on identifying stars with minimal extinction. However, the Coalsack molecular cloud lies in the Galactic plane, complicating the identification of zero-reddening stars along this sight line. 

To solve this problem, we expand our search to a more diffuse region, the intrinsic colors region (ICR), ensuring sufficient zero-reddening stars for precise $C^0_{\lambda_1,\lambda_2}$ calculations. As shown in \autoref{tab:samples}, it spans \(296^\circ \leqslant l \leqslant 312^\circ\) and \(-5^\circ \leqslant  b  \leqslant 15^\circ\), containing 368,524 stars. 
We then establish the $\mathit{T_\mathrm{eff}}$-$C^0_{\lambda_1,\lambda_2}$ relationships across 20 bands. Here is the specific procedure:
\begin{enumerate}
    \item Due to the effect of [M/H] on $C^0_{\lambda_1,\lambda_2}$ in short-wavelength bands, but minimal effect in long-wavelength bands, the optical-NIR bands ($\mathit{G_\mathrm{BP}}$, $\mathit{G_\mathrm{RP}}$,  $\mathit{B}$, $\mathit{V}$, $\mathit{u}$, $\mathit{v}$, $\mathit{g}$, $\mathit{r}$, $\mathit{i}$, $\mathit{z}$, $\mathit{J}$, $\mathit{H}$, and $\mathit{K}_\mathrm{S}$) are divided into three groups by [M/H]: \(-1\leqslant \mathrm{[M/H]} \leqslant -0.5 \) dex, \(-0.5\leqslant \mathrm{[M/H]} \leqslant 0\) dex, and \(0\leqslant \mathrm{[M/H]} \leqslant 0.5 \) dex. In contrast, the NIR-MIR bands ([3.6], [4.5], [5.8], [8.0], $\mathit{W}$1, $\mathit{W}$2, and $\mathit{W}$3) are treated as an entity range without subdivisions.
    \item The zero-reddening stars are selected in the $\mathit{T_\mathrm{eff}}$ versus Gaia ($G_\mathrm{BP}-G_\mathrm{RP}$) and $\mathit{T_\mathrm{eff}}$ versus 2MASS ($J-K_\mathrm{S}$) diagrams. For each $\mathit{T_\mathrm{eff}}$ bin of 100 K 
    moving with a window of 50 K, we select the bluest 3\% of stars. They are further required to have the G-band extinction $A_\mathrm{G}<$ 0.05 mag, as retrieved from the Gaia DR3 catalog.
    \item The relations of $\mathit{T_\mathrm{eff}}$-$C^0_\mathrm{{G_{RP}},{\lambda_1}}$ in the optical-NIR bands are derived by fitting zero-reddening stars with a cubic polynomial function. So does the relations of $\mathit{T_\mathrm{eff}}$-$C^0_\mathrm{J,{\lambda_2}}$ in the NIR-MIR bands. Here, ${\mathrm{\lambda_1}}$ denotes bands from Gaia DR3, APASS, SMSS DR4, and 2MASS, while ${\mathrm{\lambda_2}}$ includes bands from 2MASS, GLIMPSE, and WISE. The 2MASS band serves as a bridge band for subsequent analysis.
\end{enumerate}

\begin{figure} 
\centering
\includegraphics[width=1\textwidth]{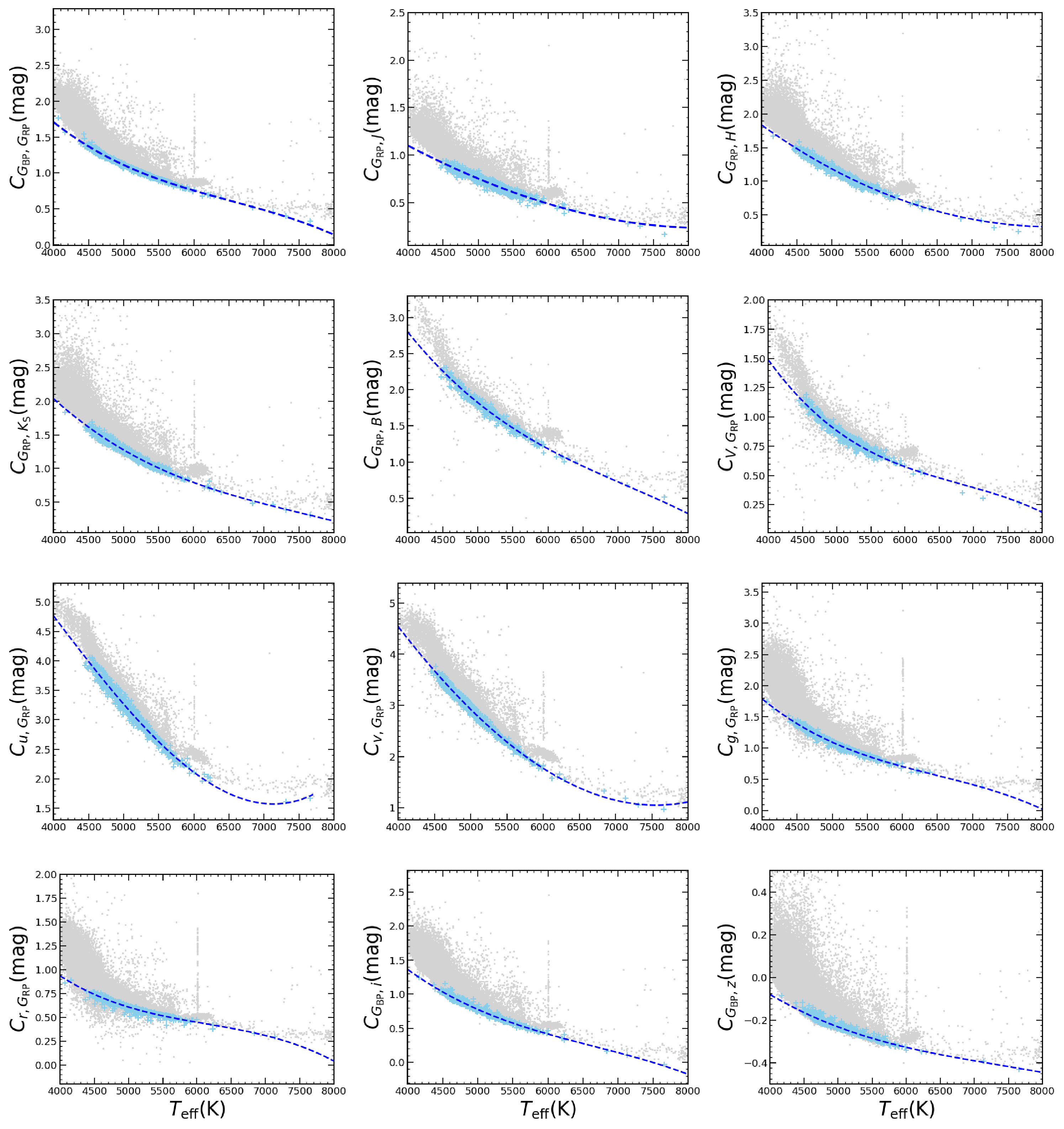}
\caption{Determination of intrinsic color index ${C^{0}_{\lambda_1,\lambda_2}}$ using the blue-edge method. Gray dots represent stars with \(-0.5\leqslant \mathrm{[M/H]} \leqslant 0\) dex in Intrinsic Colors Region (\(296^\circ \leqslant l \leqslant 312^\circ\), \(-5^\circ \leqslant  b  \leqslant 15^\circ\)) in optical-NIR bands on $\mathit{T_\mathrm{eff}}$ vs. observed color ${C^\mathrm{obs}_{\lambda_1,\lambda_2}}$ diagram. Light blue crosses denote zero-reddening stars. The blue dashed lines show the $\mathit{T_\mathrm{eff}}–{C^{0}_{\lambda_1,\lambda_2}}$ relation.}
\label{fig:blueedge}
\end{figure}

\begin{figure}
    \centering
    \includegraphics[width=0.7\linewidth]{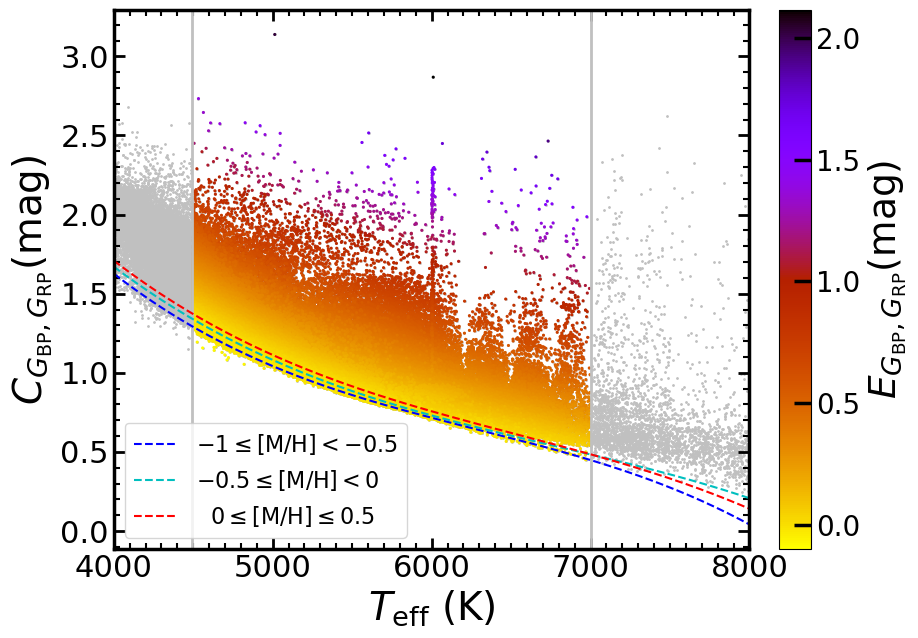}
    \caption{Observed color ${C^{\mathrm{obs}}_{\mathrm{G_{BP}, G_{RP}}}}$ vs. ${T_\mathrm{eff}}$ diagram for the Intrinsic Colors Region. Dots represent all stars, colored by their reddening $E_\mathrm{G_{BP}, G_{RP}}$. The ${T_\mathrm{eff}}$–${C^{0}_{\lambda_1,\lambda_2}}$ relations for different [M/H] intervals are shown by blue, cyan, and red dashed lines. The temperature range from 4500K to 7000K defines the reliable interval for calculating ${C^0}_{\lambda_1,\lambda_2}$.}
    \label{fig:fitline}
\end{figure}

\autoref{fig:blueedge} is an example of the measurement of the optical-NIR $C^\mathrm{0}_{\lambda_1,\lambda_2}$ for stars with \(-0.5\leqslant \mathrm{[M/H]} \leqslant 0 \) dex in the ICR.
Obviously, the fit for every band is good. However, when comparing results across different $\mathrm{[M/H]}$, the lines may cross, reverse, or diverge below 4500 K or above 7000 K, revealing slight discrepancies. As shown in \autoref{fig:fitline}, the $\mathit{T_\mathrm{eff}}$-${C^0_{\lambda_1,\lambda_2}}$ relations vary across different [M/H], i.e. the red and cyan lines coincide at 7000 K but diverge at higher $\mathit{T_\mathrm{eff}}$. To ensure accurate ${C^0_{\lambda_1,\lambda_2}}$ values across all bands, we only use the 4500 K to 7000 K range as the reliable temperature interval to calculate ${C^0_{\lambda_1,\lambda_2}}$, marked by the gray vertical lines in \autoref{fig:fitline}.

\subsection{Color Excess Ratio} \label{subsec:Color Excess Ratio}
The Color Excess (CE) is the difference between ${C^\mathrm{obs}_{\lambda_1,\lambda_2}}$ and ${C^0_{\lambda_1,\lambda_2}}$, written as $E_{\lambda_1 , \lambda_2}$ i.e. $E_{\lambda_1 , \lambda_2} = C^\mathrm{obs}_{\lambda_1,\lambda_2} - C^0_{\lambda_1,\lambda_2}$. 
\autoref{fig:fitline} illustrates the reddening values of $E_\mathrm{G_{BP}, G_{RP}}$ by the color of the dots, with closer to the blue-edge representing ${C^0_{\lambda_1,\lambda_2}}$ corresponding to smaller $E_\mathrm{G_{BP}, G_{RP}}$.

We adopt the CE method to derive the CER by calculating the ratio $k_\lambda$ of two CEs for a group of stars, utilizing the $\mathrm{CE}$-$\mathrm{CE}$ diagram \citep{2009Gao}. 
Using high-quality photometric bands as base bands for CER analysis can effectively reduce fitting errors \citep{2019wangshu}. Thus, $\mathit{G_\mathrm{BP}}$ and $\mathit{G_\mathrm{RP}}$ serve as base bands for the optical-NIR range, while $\mathit{J}$ and $K_\mathrm{S}$ are used for the NIR-MIR range. A linear fit to the $\mathrm{CE}$-$\mathrm{CE}$ diagram yields the CER
\begin{equation}
\begin{aligned}
k_{\lambda_1} &= \frac{E_\mathrm{{G_{RP},\lambda_1}}}{E_\mathrm{{G_{BP} , G_{RP}}}}=\frac{A_\mathrm{G_{RP}} - A_{\lambda_1}}{A_\mathrm{G_{BP}} - A_\mathrm{G_{RP}}} \\
k_{\lambda_2} &=  \frac{E_\mathrm{{J ,\lambda_2}}}{E_\mathrm{{J , K_S}}}=\frac{A_\mathrm{J} - A_{\lambda_2}}{A_\mathrm{J} - A_\mathrm{K_S}},
\end{aligned}
\end{equation}
where $\lambda_1$ and $\lambda_2$ refer to the same bands as defined in Section \ref{subsec:Intrinsic Color Index}.

The Coalsack molecular cloud exhibits non-uniform environmental conditions. In denser regions, dust grains tend to grow, altering size and composition, which in turn affects the infrared extinction law. To explore these variations, we define two specific regions: an inner dense region (${E_\mathrm{{G_{BP} , G_{RP}}}}>$ 1.25 mag or ${E_\mathrm{{J , K_S}}}>$ 0.5 mag) and an inner diffuse region ($E_{\mathrm{{G_{BP}, G_{RP}}}} \leq $1.25 mag or $E_{\mathrm{{J, K_S}}} \leq $0.5 mag). Dense regions typically exhibit higher extinction and larger CE values. While a single overall fit is applied to four bands ($\mathit{W}$3, $\mathit{B}$, $\mathit{u}$, $\mathit{v}$) due to the lack of high-extinction sources, all other bands are analyzed separately for the two defined specific regions. 

\begin{figure}[ht!]
\centering
\includegraphics[width=0.74\textwidth]{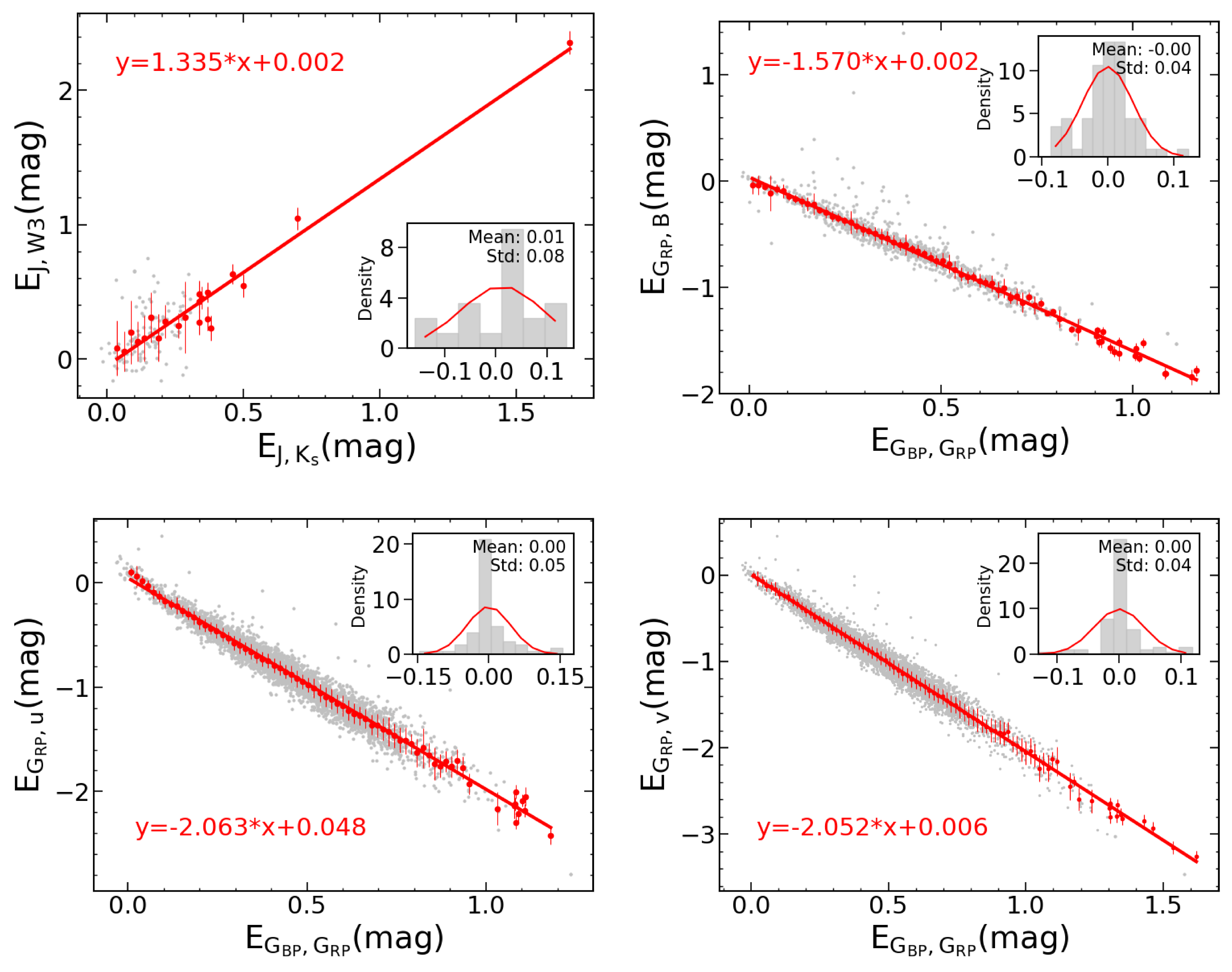}
\caption{CE-CE diagrams for the Coalsack molecular cloud in four bands ($\mathit{W}$3, $\mathit{B}$, $\mathit{u}$, and $\mathit{v}$). Red lines are the best-fit linear lines. Gray dots represent all stars, while red dots with error bars show the median values for each bin, with $E_\mathrm{{G_{BP} , G_{RP}}}$ and $E_\mathrm{{J , K_S}}$ values binned in 0.01 mag intervals. Fitting results are provided on each panel. Data distribution, residuals, standard deviations, and mean values are shown in a small inset in each corner.
\label{fig:CERS1X4}}
\end{figure}

\autoref{fig:CERS1X4} shows the CE-CE diagrams of the Coalsack region across four bands ($\mathit{W}$3, $\mathit{B}$, $\mathit{u}$, $\mathit{v}$), with gray dots representing all stars. Low-extinction stars are numerous and clustered, while high-extinction stars are sparse and scattered. This distribution inhomogeneity presents a challenge in fitting both types with minimal deviation. To overcome this, we employ the method from \cite{2021Sun}, segmenting $E_\mathrm{{G_{BP} , G_{RP}}}$ and ${E_\mathrm{{J, K_S}}}$ values into 0.01 mag bins. 
The median of each bin is calculated by iteratively clipping stars beyond $3\sigma$, with bins containing less than 10 sources excluded from the fitting. To retain information from rare but crucial high-extinction sources, all data points are kept in intervals where ${E_\mathrm{{G_{BP}, G_{RP}}}} > 1.25$ mag or ${E_\mathrm{{J, K_S}}} > 0.5$ mag, regardless of source count.
In \autoref{fig:CERS1X4}, the optimal linear fits are marked in red, and red dots with error bars signify the median values for each bin. The inset displays the mean and standard deviation of residuals, with an average standard deviation across the four bands of 0.05 $\pm$ 0.01 mag. 

\begin{figure}[ht!]
\centering
\includegraphics[width=1\textwidth]{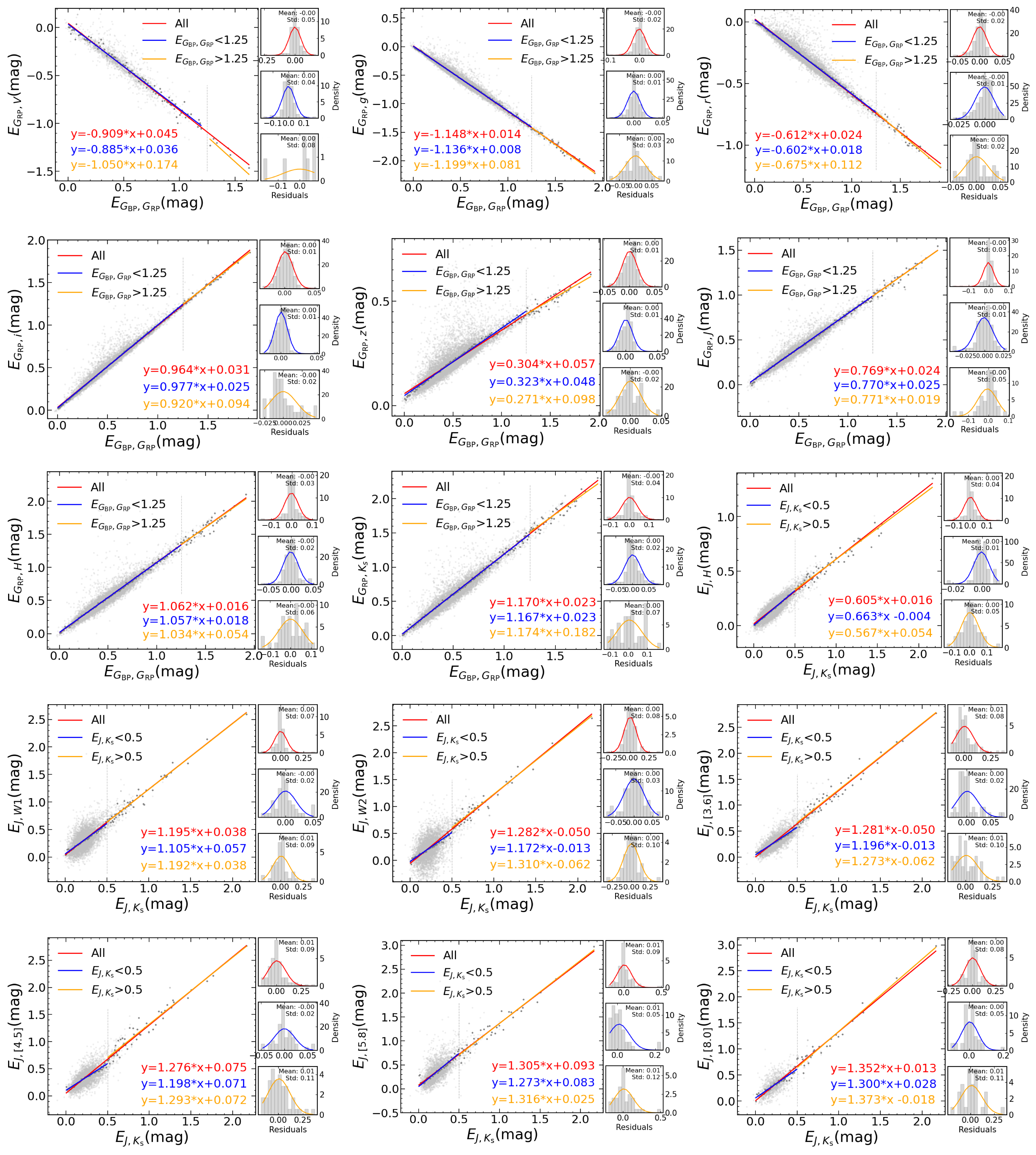}
\caption{CE-CE diagrams of $E_\mathrm{G_{RP},\lambda_1}$ vs. $E_\mathrm{G_{BP} , G_{RP}}$ and $E_\mathrm{J , \lambda_2}$ vs. $E_\mathrm{J , K_S}$. Here, ${\mathrm{\lambda_1}}$ refers to $\mathit{V}$ band from APASS, and $\mathit{g}$, $\mathit{r}$, $\mathit{i}$, $\mathit{z}$ bands from SMSS. ${\mathrm{\lambda_2}}$ includes $\mathit{J}$, $\mathit{H}$, and $\mathrm{K_S}$ bands from 2MASS; [3.6], [4.5], [5.8], [8.0] bands from GLIMPSE; and $\mathit{W}$1, $\mathit{W}$2 bands from WISE. 
The best-fitting lines for the three regions—the Coalsack region, the inner diffuse region, and the inner dense region—are shown in red, blue, and yellow, respectively. 
The gray dashed line separates the inner dense region and the inner diffuse region. 
Additionally, residual plots for each band are displayed on the right, with the CERs (slopes of the linear fits) shown in the lower corner.
\label{fig:CERS}}
\end{figure}

\autoref{fig:CERS} displays the CERs for the entire Coalsack region of the bands except those in \autoref{fig:CERS1X4}, as well as its inner dense and diffuse regions. The average standard deviation of residuals across the 15 bands is 0.04 $\pm$ 0.02 mag for the cloud, 0.02 $\pm$ 0.01 mag for the inner diffuse region, and 0.07 $\pm$ 0.02 mag for the inner dense region. 
The CERs exhibit regional variations as shown in \autoref{fig:CERS} with higher in dense regions, lower in diffuse regions, and intermediate across the entire cloud. The CERs for Ref. 1 and Ref. 2 are summarized in \autoref{tab:CERS}.

\subsection{Relative Extinction} \label{subsec:Relative Extinction}
The relative extinction for the optical-NIR bands is expressed as $A_\lambda/A_\mathrm{V}$, while for the NIR-MIR bands it is represented by $A_\lambda/A_\mathrm{{K_S}}$. These values can be converted from CERs by 
\begin{align}
& \frac{A_{\lambda_1}}{A_\mathrm{{G_{RP}}}} = 1 + k_{\lambda_1} \left( 1- \frac{A_\mathrm{{G_{BP}}}}{A_\mathrm{{G_{RP}}}} \right)  \label{eq:cer1} \\
& \frac{A_{\lambda_2}}{A_\mathrm{{K_S}}} = \frac{A_\mathrm{J}}{A_\mathrm{{K_S}}} + k_{\lambda_2} \left(1- \frac{A_\mathrm{J}}{A_\mathrm{{K_S}}} \right) , \label{eq:cer3}
\end{align}
where $A_\mathrm{G_{BP}}/A_\mathrm{G_{RP}}$ and $A_\mathrm{J}/A_\mathrm{{K_S}}$ are required for the conversion.
We adopt the $A_\mathrm{G_{BP}}/A_\mathrm{G_{RP}}$ value from the extinction law of 
\cite{2019wangshu}, hereafter \hyperref[wc19_label]{WC19}\label{wc19_label}. 

Slight differences in the effective wavelengths of different star types may lead to variations in $A_\mathrm{G_{BP}}/A_\mathrm{G_{RP}}$.To account for this, the effective wavelength of each band is calculated as
\begin{equation}
\lambda_{\mathrm{eff}, 0}=\frac{\int \lambda F_\lambda(\lambda) S(\lambda) d \lambda}{\int F_\lambda(\lambda) S(\lambda) d \lambda} ,
\label{eq:lambda}
\end{equation}
where $F_\lambda(\lambda)$ is the stellar intrinsic flux and $S(\lambda)$ is the filter transmission curve. 
We calculate the average values of $\mathit{T_\mathrm{eff}}$, $\log g$, and [M/H] for each sample
and select the corresponding synthetic stellar spectra $F_\lambda(\lambda)$ from \cite{Lejeune1997A&AS..125..229L}. The derived $\mathrm{\lambda_{eff,0}}$ values are listed in the second column of \autoref{tab:CERS}. Then $A_\mathrm{G_{BP}}/A_\mathrm{G_{RP}}$ values are obtained based on the \hyperref[wc19_label]{WC19}\label{wc19_label} extinction law with the derived $\mathrm{\lambda_{eff,0}}$. The corresponding $A_\mathrm{J}/A_\mathrm{{K_S}}$ value used in \autoref{eq:cer3} is calculated from $A_\mathrm{J}/A_\mathrm{G_{RP}}$ and $A_\mathrm{{K_S}}/A_\mathrm{G_{RP}}$, both derived from \autoref{eq:cer1}. After that, the optical-NIR relative extinction $A_\lambda/A_\mathrm{V}$ and NIR-MIR relative extinction $A_\lambda/A_\mathrm{K_S}$ are calculated from \autoref{eq:cer1} and \autoref{eq:cer3}, respectively. All results are summarized in \autoref{tab:Relative Extinction}.

\section{Results and Discussions} \label{sec:Results and Discussion}
\subsection{Reddening and Extinction Laws}
The reddening curves for the Coalsack and reference regions are analyzed using the indicator of CER $E_\mathrm{{G_{RP},\lambda}}/{E_\mathrm{{G_{BP} , G_{RP}}}}$ or $E_\mathrm{{J ,\lambda}}/E_\mathrm{{J , K_S}}$, while the extinction curves are analyzed using the indicator of relative extinction $A_\lambda/A_\mathrm{V}$ or $A_\lambda/A_\mathrm{{K_S}}$.

\subsubsection{Reddening and Extinction Law in Optical-Near-infrared}
\label{subsec:Optical-NIR}
\begin{figure}[ht!]
\centering
\includegraphics[width=0.7\textwidth]{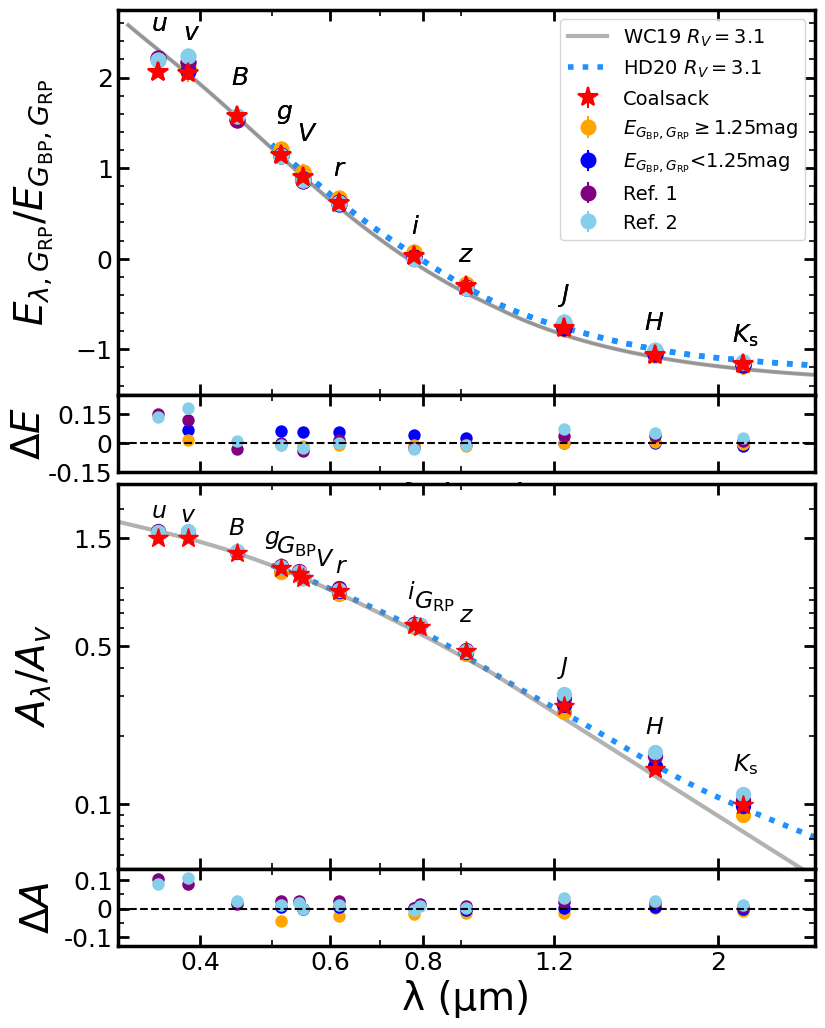}
\caption{The top panel shows the optical-NIR CERs vs. $\mathrm{\lambda_{eff}}$ diagram, while the bottom panel presents the relative extinction $A_\lambda/A_\mathrm{V}$, with differences from the Coalsack indicated by $\Delta\mathit{E}$ and $\Delta\mathit{A}$. Data points from different regions are colored. 
Yellow dots represent the inner dense region ($E_\mathrm{G_{BP}, G_{RP}} \geq 1.25$ mag), blue dots depict the inner diffuse region ($E_\mathrm{G_{BP}, G_{RP}} < 1.25$ mag), purple dots refer to Ref. 1, and light blue dots to Ref. 2.
Extinction curves from \citeauthor{2019wangshu} (2019, gray solid line) and \citeauthor{2020H&D} (2020, blue dotted line) are also shown.
\label{fig:ebprp_new}}
\end{figure}

\autoref{fig:ebprp_new} shows the optical-NIR reddening curve (top panel) with $E_\mathrm{{G_{RP},\lambda}}/{E_\mathrm{{G_{BP} , G_{RP}}}}$ and the extinction curve (bottom panel) with  $A_\lambda/A_\mathrm{V}$, where $\Delta\mathit{E}$ and $\Delta\mathit{A}$ present the difference between the reference region and the Coalsack region. For comparison, the extinction curves of \hyperref[wc19_label]{WC19} (solid gray line) and \citeauthor{2020H&D}
(2020, hereafter \hyperref[HD20_label]{HD20}\label{HD20_label}, blue dashed line) are also displayed. 

As seen in the top panel, except for the deviation in the $\mathit{u}$ and $\mathit{v}$ bands, the CERs for other bands follow the extinction law of $R_\mathrm{V}$ = 3.1. In Ref. 2, the $\mathit{v}$-band CER is slightly elevated, while other regions satisfy $R_\mathrm{V}$ = 3.1. In contrast, the $\mathit{u}$-band CERs are notably lower, especially in the Coalsack region. \cite{2019wangshu} identified a systematic curvature in $\mathit{u}$-band CER from the Sloan Digital Sky Survey, and \cite{2023Zhang&Yuan} and \cite{2023Lil} discussed CER variation with $\mathit{T_\mathrm{eff}}$. 
We also detect a slight curvature in the $\mathit{u}$ band and apply a correction based on the method of \cite{2019wangshu}, but the result changes little. Additionally, a red leak ($\sim$1\%) in the $\mathit{u}$ band (700–750 nm) \citep{2024smssdr4} may contribute to the observed deviations. So the $\mathit{u}$-band deviation may not be true.

The bottom panel presents the optical-NIR relative extinction $A_\lambda/A_\mathrm{V}$. In the densest region with $E_\mathrm{G_{BP} , G_{RP}}\geq$ 1.25 mag, the $A_\lambda/A_\mathrm{V}$ values are the lowest, while the most diffuse region (Ref. 2) shows the highest values. This suggests denser regions exhibit lower $A_\lambda/A_\mathrm{V}$ values. In the near-ultraviolet ($\mathit{u}$) and NIR ($\mathit{J}$, $\mathit{H}$, and $\mathit{K}_\mathrm{S}$) bands, relative extinction shows slight regional variation, but medians remain consistent with the $R_\mathrm{V}$ = 3.1 of \hyperref[HD20_label]{HD20}\label{HD20_label}. Other bands show little variation, consistently following $R_\mathrm{V}$ = 3.1.

\subsubsection{Reddening and Extinction Law in Near and Mid-infrared} 
\label{subsec:NIR-MIR}
\begin{figure}
    \centering
    \includegraphics[width=1\linewidth]{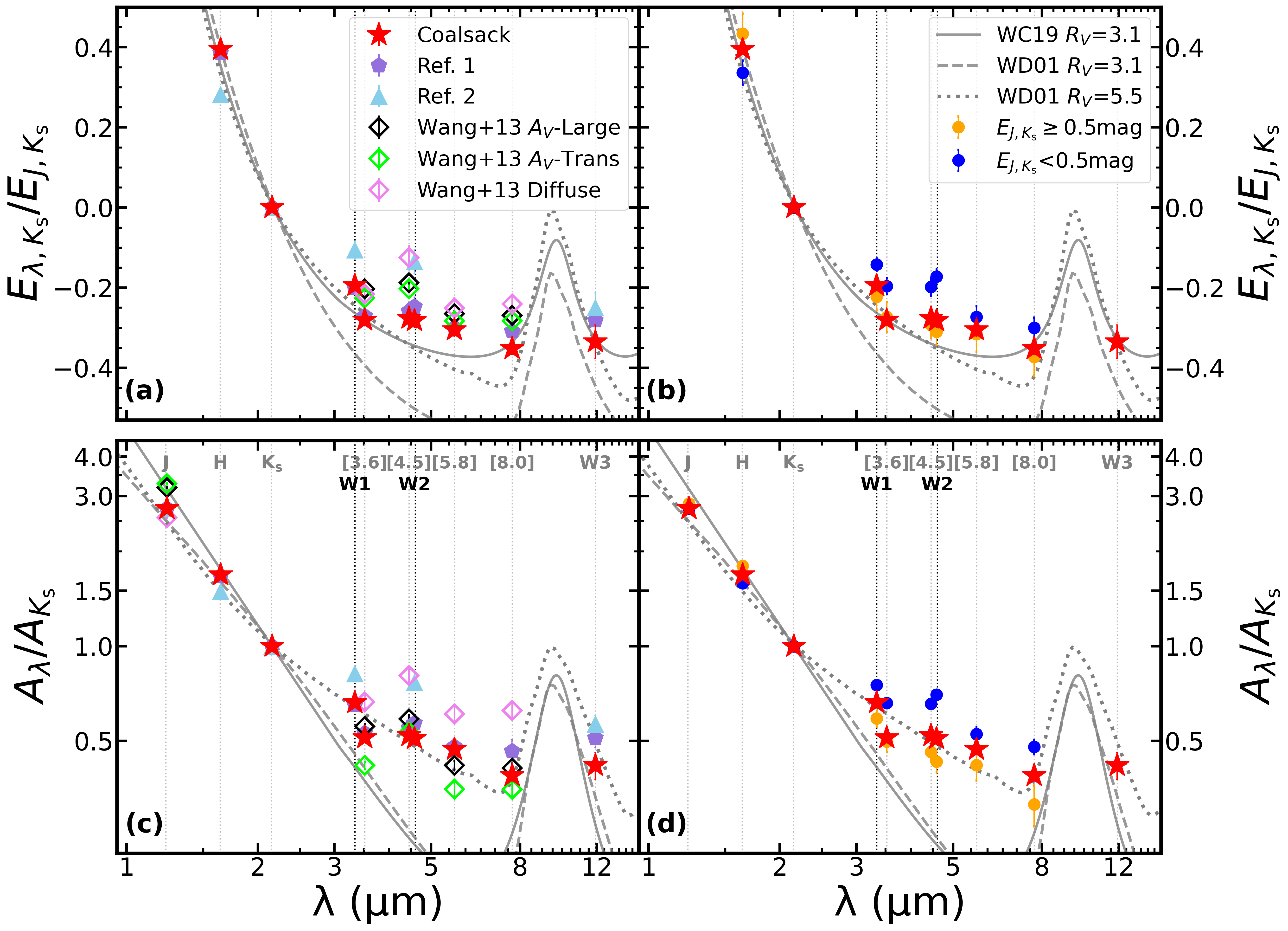}
    \caption{The NIR$-$MIR reddening curves are shown in the upper panels (a) and (b), with the CERs, while the NIR$-$MIR extinction curves are presented in the lower panels (c) and (d), with $A_\lambda/A_\mathrm{K_S}$. Different regions—Coalsack, Ref. 1, Ref. 2, inner dense, and inner diffuse—are represented by distinct symbols. For comparison, the diffuse, $A_\mathrm{V}$-Large, and $A_\mathrm{V}$-Trans regions from \cite{2013Wangshu_coalsack} are shown as open diamonds with error bars in pink, black, and green. The extinction laws from \cite{2019wangshu} and \cite{2001wd5.5} are  included. Ref. 2, with its high Galactic latitude, lacks the GLIMPSE data.}
    \label{fig:AK}
\end{figure}

\autoref{fig:AK} presents the NIR–MIR reddening curve (top panels) with $E_\mathrm{J,\lambda}/E_\mathrm{J , K_S}$ and extinction curve (bottom panels) with $A_\lambda/A_\mathrm{{K_S}}$. The left panels (a, c) compare the Coalsack region with two external reference regions (Ref. 1 and Ref. 2) and previous studies, while the right panels (b, d) focus on the comparison within the Coalsack. The figure illustrates regional variation, to varying degrees, in the NIR-MIR extinction $A_\lambda/A_\mathrm{K_S}$ across different interstellar environments. In \autoref{fig:AK} (a), the Coalsack (the densest environment) exhibits the lowest CERs, while Ref. 2 (the most diffuse) shows the highest CERs. In \autoref{fig:AK} (b), $A_\lambda/A_\mathrm{K_S}$ values are the lowest in the inner dense region and the highest in the inner diffuse region. 

\cite{2013Wangshu_coalsack} identified three regions—diffuse, $A_\mathrm{V}$-Large, and $A_\mathrm{V}$-Trans—using \cite{2005Dobashi}’s $A_\mathrm{V}$ map, the Spitzer/MIPS 24 $\mu$m emission map, and CO gas emission contours. 
By comparison, we find that their diffuse region, near the eastern cloud edge with low dust and CO emission, corresponds to our Ref. 2. Their $A_\mathrm{V}$-Large aligns with our Coalsack region, and $A_\mathrm{V}$-Trans similar to our Ref. 1 region.
In \autoref{fig:AK} (a) and (c), Ref. 2 displays higher values, suggesting unique extinction properties resembling \cite{2013Wangshu_coalsack}’s diffuse region, with both being flatter than others, even more than \hyperref[WD01_label]{WD01} $R_\mathrm{V}$ = 5.5. In contrast, CERs in the Coalsack and Ref. 1 are lower than those in $A_\mathrm{V}$-Large and $A_\mathrm{V}$-Trans. For $A_\lambda/A_\mathrm{K_S}$ values, the Coalsack and Ref. 1 closely match $A_\mathrm{V}$-Large, while $A_\mathrm{V}$-Trans values remain lower. 
Although our results are generally consistent with \cite{2013Wangshu_coalsack}, we observe lower CERs values and slight difference in relative extinction, likely due to methodological differences. Unlike \cite{2013Wangshu_coalsack}, which sampled sub-regions without correcting for $C^0_{\lambda_1,\lambda_2}$, we examine the entire cloud, accurately defined, using high-quality photometric data and $C^0_{\lambda_1,\lambda_2}$ corrections. 
In addition, \cite{2013Wangshu_coalsack} used photometric data which may probe deeper region than this work. Nevertheless, this broader approach enhances our findings and refines established extinction laws.
Overall, the Coalsack extinction law shows a steep slope in the NIR bands, satisfying the extinction law of $R_\mathrm{V}$ = 3.1, and flattens in the MIR bands, following \hyperref[WD01_label]{WD01} $R_\mathrm{V}$ = 5.5. As a quiescent, starless molecular cloud, the Coalsack’s optical–MIR extinction law aligns with those of the active star-forming clouds \citep{2023Li,2023Cao,2024Li}.

\subsection{The Reddening Map of $E_\mathrm{B,V}$}
\begin{figure}
    \centering
    \includegraphics[width=0.8\linewidth]{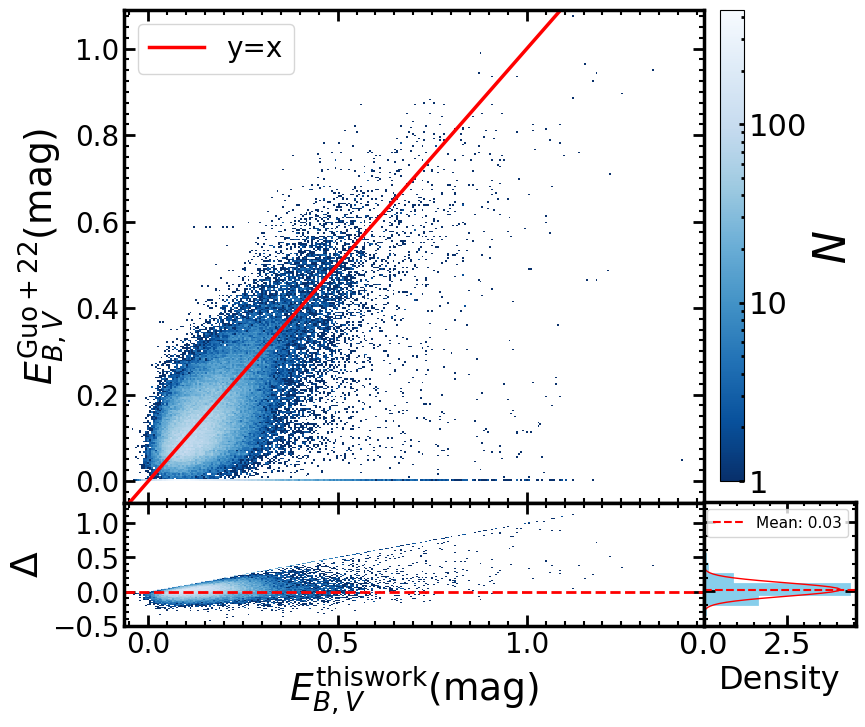}
    \caption{Top panel: $E_\mathrm{B,V}$ diagram for 128,645 stars in the region \(296^\circ \leqslant l \leqslant 312^\circ\) , \(-5^\circ \leqslant  b  \leqslant 5^\circ\), colored by star number density. The x-axis shows our $E_\mathrm{B,V}$ values ($E_\mathrm{B,V}^\mathrm{this work}$), while the y-axis shows those from \citeauthor{2022Guomap} (2022, $E_\mathrm{B,V}^\mathrm{Guo+22}$), with a red solid line indicating the y = x relation. Bottom panel: the distribution of residuals ($\Delta = E_\mathrm{B,V}^\mathrm{this work} - E_\mathrm{B,V}^\mathrm{Guo+22}$) is shown on the left, with a red dashed line at y = 0. A histogram of residuals is presented on the right. The solid red curve is a Gaussian fit to the distribution of differences, and a red dashed line marks the mean.}
    \label{fig:guo1}
\end{figure}

\begin{figure}
    \centering
    \includegraphics[width=0.9\linewidth]{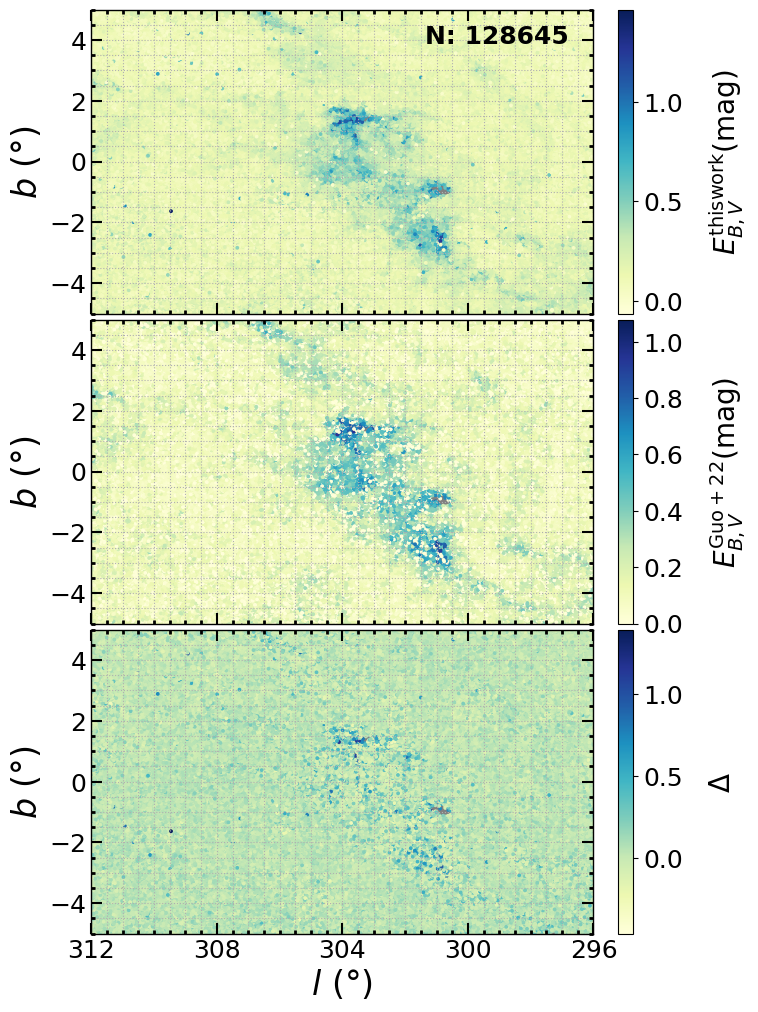}
    \caption{Spatial distributions of $E_\mathrm{B,V}$ values from this work $E_\mathrm{B,V}^\mathrm{this work}$ and \citeauthor{2022Guomap} (2022, $E_\mathrm{B,V}^\mathrm{Guo+22}$) with the residuals ($\Delta = E_\mathrm{B,V}^\mathrm{this work} - E_\mathrm{B,V}^\mathrm{Guo+22}$). The number of sources is indicated in bold black text in the top-right corner. 
    Core regions with high extinction, such as $\mathit{l}$= 300.6° and $\mathit{b}$ = -1°, lack CE availability due to data limitations.}
    \label{fig:guo2}
\end{figure}

\cite{2022Guomap} presented a catalog of molecular clouds in the Southern sky, including the Coalsack, with $E_\mathrm{B,V}$ values. Our $E_\mathrm{B,V}$ measurement  ($E_\mathrm{B,V}^\mathrm{this work}$) in the region \(296^\circ \leqslant l \leqslant 312^\circ\), \(-5^\circ \leqslant  b  \leqslant 5^\circ\) are compared with those from \cite{2022Guomap}, $E_\mathrm{B,V}^\mathrm{Guo+22}$ within 1 kpc in \autoref{fig:guo1}. Good agreement is found, with a mean residual of 0.03 mag. 
In high-extinction regions, higher values are observed with a maximum $E_\mathrm{B,V}^\mathrm{thiswork}$ of 1.47 mag, exceeding the 1.08 mag in $E_\mathrm{B,V}^\mathrm{Guo+22}$.
There are 13,003 sources, i.e. 10\% of the total sample, 
with $E_\mathrm{B,V}^\mathrm{Guo+22}$ = 0 mag, but $E_\mathrm{B,V}^\mathrm{thiswork}$ $\neq$ 0 mag with an average of 0.2 mag and 98\% from 0 to 0.3 mag. 
For these sources, $E_\mathrm{B,V}^\mathrm{this work}$ values are in excellent agreement with those from \cite{2024zhoahe2} with mean difference of -0.01 mag and Gaia DR3 with mean difference of -0.02 mag. In addition, the 13,003 sources above are evenly distributed across the region without apparent clustering, while the high-extinction sources with $E_\mathrm{B,V}^\mathrm{this work} > E_\mathrm{B,V}^\mathrm{Guo+22}$ are mostly located within the cloud, as shown in \autoref{fig:guo2}.

\autoref{fig:guo2} compares the spatial distribution of $E_\mathrm{B,V}$ in this work (top panel) with that of \citeauthor{2022Guomap} (2022, middle panel), both at an angular resolution of $1.3'$.
The high-extinction regions within the Coalsack are clearly delineated, and both maps show consistent cloud boundaries. However, our result reveals finer structural details of the Coalsack, visible in the residual distribution (bottom panel). The residuals between $E_\mathrm{B,V}^\mathrm{this work}$ and $E_\mathrm{B,V}^\mathrm{Guo+22}$ are low ($\sim$ 0.01 mag) for $E_\mathrm{B,V}^\mathrm{this work} < 0.4$ mag but increase to 0.15 mag for higher values, indicating larger discrepancies at higher $E_\mathrm{B,V}$.

To summarize, the difference likely results from the methodological variation, with \cite{2022Guomap} identifying significantly more zero-extinction sources than we do. Moreover, the use of different photometric datasets may also contribute. \cite{2022Guomap} used SMSS DR1 and Gaia DR2, while we rely on SMSS DR4 and Gaia DR3, which offer improved photometric quality. This enables more accurate extinction measurements through precise $C^0_{\lambda_1,\lambda_2}$ calculations, revealing finer cloud structure and laying the groundwork for future investigations of the cloud and its surrounding environment.

\subsection{Comparison of $A_\mathrm{V}$} \label{subsec:Extinction Map}
\begin{figure}
    \centering
    \includegraphics[width=0.7\linewidth]{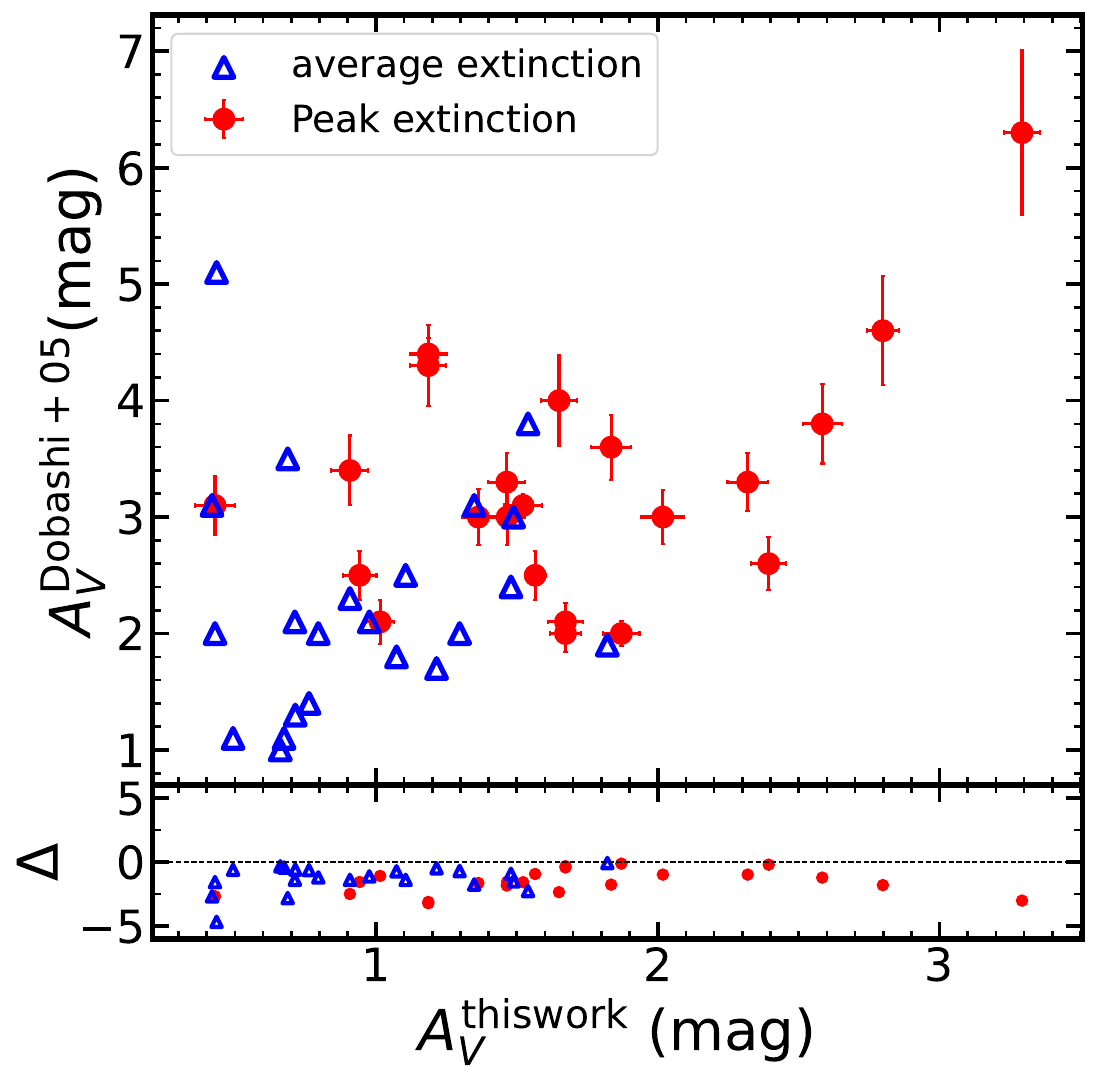}
    \caption{Comparison of $A_\mathrm{V}$ values between this work ($A_\mathrm{V}^\mathrm{this work}$) and \citeauthor{2005Dobashi} (2005, $A_\mathrm{V}^\mathrm{Dobashi+05}$) for the same sub-regions within the Coalsack cloud.
    Red dots indicate the peak $A_\mathrm{V}$ in each region, with vertical error bars indicating the uncertainties in \cite{2005Dobashi}'s peak measurements, and horizontal error bars representing the uncertainties in our peak measurements. Blue triangles represent the average $A_\mathrm{V}$ for each region. The lower panel shows the residuals ($\Delta = A_\mathrm{V}^\mathrm{this work} - A_\mathrm{V}^\mathrm{Dobashi+05}$), with a black dashed line indicating zero residual.}
    \label{fig:DOBASHI}
    
\end{figure}

\cite{2005Dobashi} presented an $A_\mathrm{V}$ map of the Coalsack molecular cloud at an angular resolution of $6'$.
They also provided $A_\mathrm{V}$ measurements for 22 specific sub-regions, including the mean $A_\mathrm{V,mean}^\mathrm{Dobashi+05}$ and the peak $A_\mathrm{V,peak}^\mathrm{Dobashi+05}$ values with associated errors. 
Since the optical-NIR extinction law for the Coalsack region conforms to $R_\mathrm{V}$ = 3.1, we can directly obtain the $\mathit{V}$-band extinction value ($A_\mathrm{V}$) using the relation $R_\mathrm{V} = A_\mathrm{V}/E(B-V)$.
We calculate the the mean $A_\mathrm{V}^\mathrm{this\ work}$ values and peak $A_\mathrm{V}^\mathrm{this\ work}$ values based on individual stars from our sample located within the sub-regions defined by \cite{2005Dobashi}. \autoref{fig:DOBASHI} shows the comparison.
The $A_\mathrm{V,mean}^\mathrm{this work}$ values predominantly range from 0.5 to 1.5 mag, all below 1.9 mag. In contrast，the $A_\mathrm{V,mean}^\mathrm{Dobashi+05}$ values range from 1 – 3 mag and cluster around 2 mag, with a maximum of 5.1 mag. 

The $A_\mathrm{V,peak}^\mathrm{this work}$ span from 1 to 2 mag, reaching a maximum of 3.3 mag, while the $A_\mathrm{V,peaks}^\mathrm{Dobashi+05}$ range from 2 to 4 mag with a maximum of 6 mag. Additionally, the $A_\mathrm{V,peak}^\mathrm{this work}$ exhibit smaller error. 
The maximum difference is 5 mag between the $A_\mathrm{V,mean}^\mathrm{this work}$ and $A_\mathrm{V,mean}^\mathrm{Dobashi+05}$, 3 mag between the $A_\mathrm{V,peaks}^\mathrm{this work}$ and $A_\mathrm{V,peaks}^\mathrm{Dobashi+05}$, resulting in a residual deviation of 1.5 mag.
In general, their values are higher than ours, likely due to difference in distance of tracers. We select stars within 1 kpc as extinction tracers to avoid contamination from background clouds, while they choose all photometrically detected stars in given slight lines. For example, in the region with the largest mean difference, the $A_\mathrm{V,mean}^\mathrm{this work}$ is 0.2 mag, while the $A_\mathrm{V,mean}^\mathrm{Dobashi+05}$ is 5.1 mag, 
suggesting their detection of distant stars may overlap with other clouds behind the Coalsack, or the detection of stars embedded in very dense regions of the cloud.

\subsection{Variation of Extinction Law}
\label{subsec:RV}
\begin{figure}
    \centering
    \includegraphics[width=0.9\linewidth]{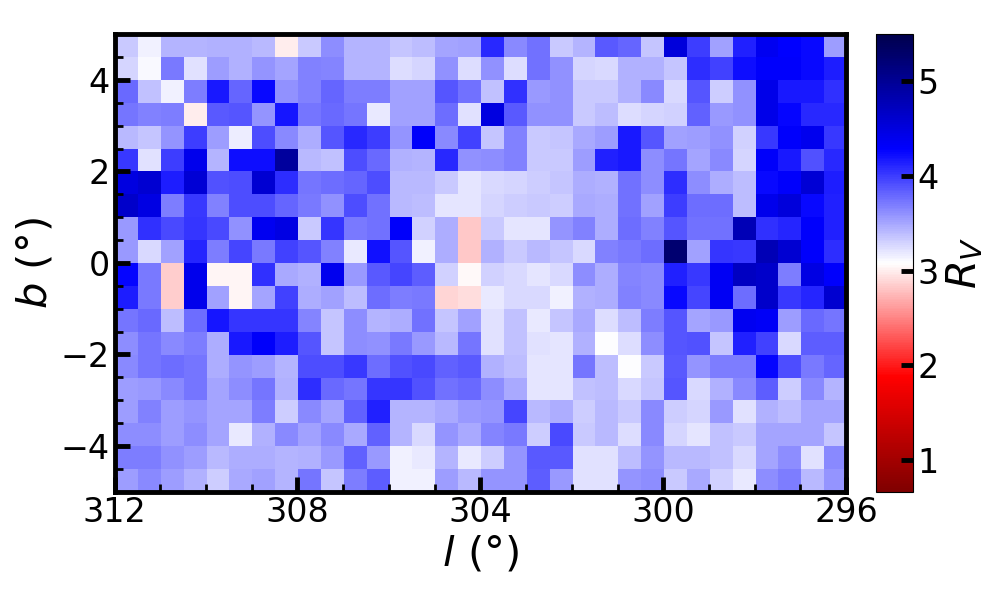}
    \caption{Spatial variation of the extinction law in the Coalsack molecular cloud within the region \(296^\circ \leqslant l \leqslant 312^\circ\), \(-5^\circ \leqslant  b  \leqslant 5^\circ\), characterized by $R_\mathrm{V}$. The region is divided into 0.5°$\times$0.5° sub-regions based on a grid of Galactic coordinates ($\mathit{l}$, $\mathit{b}$).}
    \label{fig:RV}
\end{figure}

The potential variation in extinction law for the Coalsack cloud across \(296^\circ \leqslant l \leqslant 312^\circ\) , \(-5^\circ \leqslant  b  \leqslant 5^\circ\) is probed. 
The region is first divided into 0.5°$\times$0.5° sub-regions by Galactic coordinates ($\mathit{l}$, $\mathit{b}$).
The conversion from CER to $R_\mathrm{V}$ is performed using the methods of \cite{2016Schlafly} and \cite{2023Lil}, along with the
$R_\mathrm{V}$-dependent extinction curves from \cite{2019wangshu, 2023wangshu}. To account for stellar parameter dependencies, sources are grouped into nine categories based on [M/H] with an interval of 0.5 dex, and $T_{\mathrm{eff}}$ with an interval of 1,000 K, establishing relations of $R_\mathrm{V}$-CER for each group.
This process yields nine equations to convert ${E_\mathrm{{G_{RP},\mathrm{K_S}}}}/{E_\mathrm{{G_{BP}, G_{RP}}}}$ into $R_\mathrm{V}$.
Each sub-region’s median $R_\mathrm{V}$ then characterizes its overall feature.
Spatial correlations of $R_\mathrm{V}$-CE are shown in \autoref{fig:RV}. Comparing the spatial distribution of $R_\mathrm{V}$ with CER in the top panel of \autoref{fig:guo2}, we observe a weak trend. $R_\mathrm{V}$ tends to be smaller in large $E_\mathrm{B,V}$ regions and bigger in small ones.
Previous studies yield controversial findings on $R_\mathrm{V}$-CE ($E_\mathrm{B,V}$) correlation. \cite{2016Schlafly}, \cite{2023Li}, and \cite{2024Butler} found no significant relation, while \cite{2023Zhang&Yuan} noted an inverse correlation for $E_\mathrm{B,V} < 0.1$mag. \cite{2024Zhang&Green} found a U-shaped relationship. This work further explores the weak correlation between $R_\mathrm{V}$ and CE.

\begin{figure}
    \centering
    \includegraphics[width=0.92\linewidth]{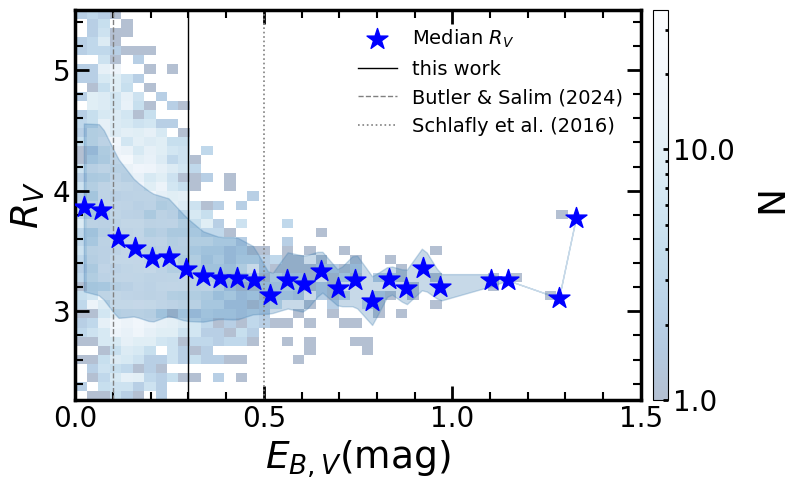}
    \caption{Distribution of $R_\mathrm{V}$ and $E_\mathrm{B,V}$ in each sub-region of \autoref{fig:RV}, with color representing the number density. Blue stars represent the median $R_\mathrm{V}$ values for each $E_\mathrm{B,V}$ bin of 
    0.05 mag, while the light blue shadow area denotes the 1$\sigma$ range. 
    The black solid line, gray dashed line, and gray dotted line indicate the minimum $E_\mathrm{B,V}$ values corresponding to the $R_\mathrm{V}$ uncertainty region identified in this work, \cite{2024Butler}, and \cite{2016Schlafly} respectively.}
    \label{fig:RV3}
\end{figure}

\begin{figure}
    \centering
    \includegraphics[width=0.63\linewidth]{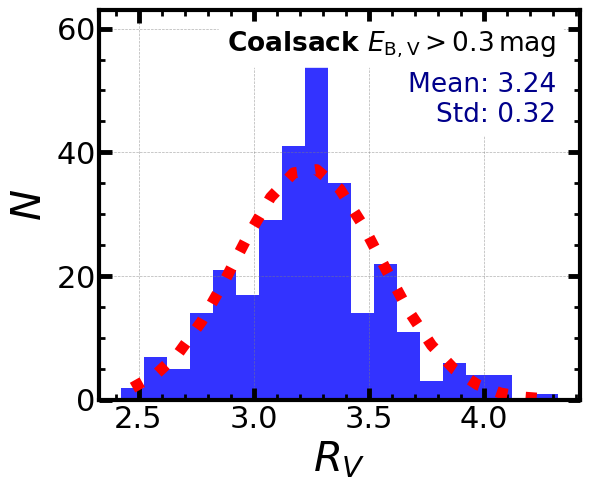}
    \caption{Distribution of $R_\mathrm{V}$ in the Coalsack molecular cloud sub-region follows a Gaussian function, with a mean value of $R_\mathrm{V} = 3.24 \pm 0.32$ for $E_\mathrm{B,V} > 0.3$ mag.}
    \label{fig:RV4}
\end{figure}

\autoref{fig:RV3} illustrates the $R_\mathrm{V}$-$E_\mathrm{B,V}$ relation, using each sub-region's median $E_\mathrm{B,V}$ from \autoref{fig:RV}. The figure shows a sharp decrease in $R_\mathrm{V}$ for $E_\mathrm{B,V}<$ 0.3 mag, where uncertainty is high. 
For $E_\mathrm{B,V}>$ 0.3 mag, $R_\mathrm{V}$ remains nearly flat with little dependence on $E_\mathrm{B,V}$, as described by a Gaussian function in \autoref{fig:RV4}, but increases sharply at the tail ($E_\mathrm{B,V}>$ 1.3 mag). 
This trend is consistent with previous work. \cite{2016Schlafly} noted significant uncertainty below $E_\mathrm{B,V}$ = 0.5 mag, a slight $R_\mathrm{V}$ rise at $E_\mathrm{B,V}$ = 1 mag, and a steep increase at $E_\mathrm{B,V}$ = 2 mag. Similarly, \cite{2024Butler} found increased $R_\mathrm{V}$ below $E_\mathrm{B,V} \approx 0.1$ mag and a more pronounced rise at 
$E_\mathrm{B,V}$= 1 mag. In \autoref{fig:RV3}, the solid black line marks the boundary of the low-extinction high-uncertainty range, while the dashed and dotted lines represent the boundaries from other studies.
To the left of the solid black line, the variation in $R_\mathrm{V}$ appears linked to the weak structural correlation seen in \autoref{fig:RV}. Fluctuation in $R_\mathrm{V}$ at the regions with $E_\mathrm{B,V}<$ 0.3 mag or $E_\mathrm{B,V}>$ 1.0 mag may reflect complex and uncertain influences on dust properties, including the formation of ice mantles on dust grains \citep{1988Whittet}, the accretion of metals from the gas phase onto dust \citep{2012MNRAS.422.1263H}, PAH growth \citep{2024zhangxiangyuPHA}.
Alongside these uncertainties, systematic biases—such as large $R_\mathrm{V}$ dispersion in low-extinction region and the limited data at high extinction—further reduce the correlation’s reliability.

In summary, no strong evidence supports a definitive correlation between $R_\mathrm{V}$ and $E_\mathrm{B,V}$, consistent with the findings of \cite{2016Schlafly}, \cite{2023wangshu}, \cite{2023Li}, and \cite{2024Butler}, suggesting a relatively uniform optical extinction law across the Coalsack region. 
The average extinction law for the Coalsack region targets with $E_\mathrm{B,V} > 0.3$ mag yields $R_\mathrm{V} = 3.24 \pm 0.32$, closely matching the $R_\mathrm{V} = 3.07 \pm 0.23$ from \cite{2024Zhang&Green} for the same region.

\section{Summary} 
\label{sec:Summary}
We select 368,524 dwarf stars from Gaia DR3 as extinction tracers to probe the extinction law in the Coalsack, a quiescent molecular cloud, using spectroscopic and photometric data. 
The main results are as follows:
\begin{enumerate}
    \item 
    The $\mathit{T_\mathrm{eff}}$-$C^0_{\lambda_1,\lambda_2}$ relations for dwarf stars across optical-NIR and NIR-MIR bands are established by using Gaia stellar parameters with $\mathit{T_\mathrm{eff}}$ 4500$-$7000 K.
    The color excesses are calculated for 20 bands, including the Gaia ($\mathit{G_\mathrm{BP}}$, $\mathit{G_\mathrm{RP}}$), APASS ($\mathit{B}$, $\mathit{V}$), SMSS ($\mathit{u}$, $\mathit{v}$, $\mathit{g}$, $\mathit{r}$, $\mathit{i}$, $\mathit{z}$), 2MASS ($\mathit{J}$, $\mathit{H}$, $\mathit{K}_\mathrm{S}$), WISE ($\mathit{W}$1, $\mathit{W}$2, $\mathit{W}$3), and GLIMPSE ([3.6], [4.5], [5.8], [8.0]) bands.
    Then color excess ratios for the Coalsack region and four reference regions (the surrounding region, the nearby high Galactic latitude region, the inner dense region, and the inner diffuse region) are derived and listed in \autoref{tab:CERS}. The $A_\lambda/A_\mathrm{V}$ across 13 optical-NIR bands and $A_\lambda/A_\mathrm{{K_S}}$ across 10 NIR-MIR bands are determined (\autoref{tab:Relative Extinction}). In addition, we present the extinction map for the Coalsack molecular cloud (\(296^\circ \leqslant l \leqslant 312^\circ\), \(-5^\circ \leqslant  b  \leqslant 5^\circ\)) at an angular resolution of ${\,1.3'}$.
    
    \item 
    The reddening curves and extinction curves for the Coalsack and reference regions across optical-MIR bands are derived.
    In the optical-NIR bands of 0.35–2.15 $\mu$m, our curve closely follows the Galactic average extinction law of $R_\mathrm{V}$ = 3.1. It steepens in the NIR and flattens in the MIR (2.15–12.00 $\mu$m), aligning with the \hyperref[WD01_label]{WD01} $R_\mathrm{V}$ = 5.5 curve. 
    Regional variations in the NIR-MIR extinction law are observed, with flatter profiles in diffuse regions. As a typical quiescent starless molecular cloud, the optical-MIR extinction law of the Coalsack is consistent with those found in the active star-forming molecular clouds.

    \item  
    The variation in extinction law for the Coalsack cloud (\(296^\circ \leqslant l \leqslant 312^\circ\) , \(-5^\circ \leqslant  b  \leqslant 5^\circ\)) reveals a weak structural correlation where $R_\mathrm{V}$ tends to be lower in large $E_\mathrm{B,V}$ regions and higher in small ones. 
    We further investigate the dependence of $R_\mathrm{V}$ on $E_\mathrm{B,V}$ along the line of sight.
    However, no strong correlation is found, suggesting a relatively uniform optical extinction law in the Coalsack region. 
    For $E_\mathrm{B,V}>$ 0.3 mag, the average extinction law across the Coalsack cloud is depicted by $R_\mathrm{V}$=3.24 ± 0.32.
    
\end{enumerate}
We thank Profs. Jian Gao and Haibo Yuan for the very helpful discussions and suggestion,  
and the anonymous referee for his/her helpful comments/suggestions.
This work is supported by the National Natural Science Foundation of China (NSFC) project 12133002 and 12373028, the China Manned Space Project with No. CMS-CSST-2021-A09. S.W. acknowledges the support from the Youth Innovation Promotion Association of the CAS (grant No. 2023065). This work has made use of the data from Gaia, APASS, SMSS, 2MASS, GLIMPSE, WISE surveys.

\bibliography{reference}
\bibliographystyle{aasjournal}

\newpage

\begin{table}
\centering
\caption{Different Samples Used in This Work\label{tab:samples}}
\setlength{\tabcolsep}{25pt} 
\begin{tabular}{cccc}
\toprule
\midrule 
Sample Name   &    $ l \ (^\circ)$ & $b\ (^\circ)$    & Number of Sources\\
\midrule
Coalsack             & 299～305.5     & -2.8～2.45        & 14,112                   \\
Reference region 1 (Ref. 1)         & 296～312       & -5～5             & 117,585                  \\
Reference region 2 (Ref. 2)         & 296～305       & 5～10             & 32,964                   \\
Intrinsic colors region (ICR)    & 296～312       & -5～15            & 368,524                  \\ 
\bottomrule
\end{tabular}
\end{table}


\begin{table}[b]
\rotatebox{90}{%
\begin{minipage}{\textheight} 
\centering
\caption{Multi-wavelengt Color Excess Ratios and Sample Size} 
\label{tab:CERS}
\renewcommand{\arraystretch}{1.4} 
\setlength{\tabcolsep}{2.8pt} 
\begin{tabular}{@{}cccccccccccccc@{}}
\toprule
\midrule

Band                     & $\lambda_{\text{eff}, 0}$ &  & \multicolumn{5}{c}{$E_{G_\mathrm{RP},\lambda}  /E_{G_\mathrm{BP},G_\mathrm{RP}}$}                                                                            &  & \multicolumn{5}{c}{Number of Sources}                                                                                                                                 
\\
\cline{4-8}
\cline{10-14}
                         & ($\mu$m)    &  & Coalsack                &  $E_{G_\mathrm{BP},G_\mathrm{RP}} < 1.25$  & $E_{G_\mathrm{BP},G_\mathrm{RP}}\geq1.25$ & Ref. 1               & Ref. 2               &  & Coalsack     & $E_{G_\mathrm{BP},G_\mathrm{RP}}$$<1.25$ & $E_{G_\mathrm{BP},G_\mathrm{RP}}\geq1.25$  & Ref. 1     & Ref. 2    \\
\cline{1-14}
$G_\mathrm{BP}$ & 0.5441  &  & -1               & ···                                                                 & ···                                                                  & -1               & -1               &  & 12414 & ···                                                                 & ···                                                                  & 111258 & 31651 \\
$G_\mathrm{RP}$ & 0.7927  &  & 0                & ···                                                                 & ···                                                                  & 0                & 0                &  & 13302 & ···                                                                 & ···                                                                  & 112910 & 31971 \\
$\mathit{B}$                        & 0.4487  &  & $-1.570\pm0.012$ & ···                                                                 & ···                                                                  & $-1.536\pm0.004$ & $-1.577\pm0.010$ &  & 1053  & ···                                                                 & ···                                                                  & 14460  & 8038  \\
$\mathit{V}$                        & 0.5519  &  & $-0.909\pm0.008$ & $-0.885\pm0.010$                                                    & $-1.050\pm0.080$                                                     & $-0.854\pm0.003$ & $-0.870\pm0.011$ &  & 1560  & 1553                                                                & 10                                                                   & 15771  & 9147  \\
$\mathit{J}$                        & 1.2373  &  & $0.769\pm0.003$  & $0.770\pm0.002$                                                     & $0.771\pm0.029$                                                      & $0.733\pm0.002$  & $0.700\pm0.003$  &  & 7577  & 7474                                                                & 82                                                                   & 68526  & 20791 \\
$\mathit{H}$                        & 1.6442  &  & $1.062\pm0.004$  & $1.057\pm0.003$                                                     & $1.034\pm0.035$                                                      & $1.024\pm0.002$  & $1.006\pm0.010$  &  & 8552  & 8304                                                                & 89                                                                   & 78215  & 25504 \\
$K_\mathrm{S}$           & 2.1567  &  & $1.165\pm0.005$  & $1.170\pm0.005$                                                     & $1.174\pm0.041$                                                      & $1.158\pm0.003$  & $1.141\pm0.012$  &  & 8414  & 8184                                                                & 90                                                                   & 73936  & 24017 \\
$\mathit{u}$                        & 0.3517  &  & $-2.063\pm0.008$ & ···                                                                 & ···                                                                  & $-2.212\pm0.007$ & $-2.198\pm0.011$ &  & 4500  & ···                                                                 & ···                                                                  & 59715  & 20404 \\
$\mathit{v}$                        & 0.3854  &  & $-2.052\pm0.009$ & ···                                                                 & ···                                                                  & $-2.168\pm0.005$ & $-2.232\pm0.014$ &  & 6780  & ···                                                                 & ···                                                                  & 71557  & 23462 \\
$\mathit{g}$                        & 0.5148  &  & $-1.148\pm0.002$ & $-1.136\pm0.002$                                                    & $-1.199\pm0.030$                                                     & $-1.147\pm0.002$ & $-1.134\pm0.004$ &  & 11602 & 10933                                                               & 91                                                                   & 95726  & 28016 \\
$\mathit{r}$                        & 0.6160  &  & $-0.612\pm0.001$ & $0.602\pm0.001$                                                     & $-0.675\pm0.025$                                                     & $-0.625\pm0.002$ & $-0.610\pm0.005$ &  & 11613 & 10902                                                               & 99                                                                   & 95888  & 27479 \\
$\mathit{i}$                        & 0.7767  &  & $-0.036\pm0.001$ & $-0.023\pm0.001$                                                    & $-0.080\pm0.029$                                                     & $-0.004\pm0.003$ & $0.001\pm0.004$  &  & 11652 & 10831                                                               & 96                                                                   & 94969  & 26524 \\
$\mathit{z}$                        & 0.9135  &  & $0.304\pm0.001$  & $0.323\pm0.002$                                                     & $0.271\pm0.017$                                                      & $0.312\pm0.002$  & $0.319\pm0.004$  &  & 11906 & 11084                                                               & 97                                                                   & 96938  & 26961 \\
\cline{1-14}

                   &  &  & \multicolumn{5}{c}
                      {$E_{J,\lambda} / E_{J,K_s}$}                                                                                                                           &  & \multicolumn{5}{c}{Number of Sources}                                                                                                                                  \\
\cline{4-8}
\cline{10-14}

                         &         & & Coalsack                & $ E_{J,K_s}$$<0.5$                                 & $E_{J,K_s}\geq0.5$                                  & Ref. 1               & Ref. 2               &  & Coalsack     & $E_{J,K_s}$$<0.5$                                  & $E_{J,K_s}\geq0.5$                                  & Ref. 1     & Ref. 2    \\
\cline{1-14}
$\mathit{J}$                        & 1.2373  &  & 0                & 0                                                                   & 0                                                                    & 0                & 0                &  & 7577  & ···                                                                 & ···                                                                  & 68526  & 20791 \\
$\mathit{H}$                       & 1.6442  &  & $0.605\pm0.006$  & $0.663\pm0.030$                                                     & $0.566\pm0.050$                                                      & $0.614\pm0.007$  & $0.718\pm0.011$  &  & 8552  & 3996                                                                & 144                                                                  & 78215  & 25504 \\
${K_S}$           & 2.1567  &  & 1                & 1                                                                   & 1                                                                    & 1                & 1                &  & 8414  & ···                                                                 & ···                                                                  & 73936  & 24017 \\
$\mathit{W1}$                       & 3.3397  &  & $1.211\pm0.013$  & $1.142\pm0.016$                                                     & $1.223\pm0.035$                                                      & $1.201\pm0.010$  & $1.106\pm0.022$  &  & 5241  & 4085                                                                & 149                                                                  & 60774  & 27886 \\
$\mathit{W2}$                       & 4.5811  &  & $1.282\pm0.017$  & $1.172\pm0.018$                                                     & $1.310\pm0.032$                                                      & $1.248\pm0.020$  & $1.135\pm0.025$  &  & 5090  & 4024                                                                & 148                                                                  & 60110  & 27872 \\
$\mathit{W3}$                       & 11.9151 &  & $1.335\pm0.030$  & ···                                                                 & ···                                                                  & $1.281\pm0.031$  & $1.250\pm0.040$  &  & 229   & ···                                                                 & ···                                                                  & 2203   & 967   \\
{[}3.6{]}                & 3.5205  &  & $1.281\pm0.013$  & $1.196\pm0.031$                                                     & $1.273\pm0.050$                                                      & $1.271\pm0.019$  & ···              &  & 6377  & 3811                                                                & 138                                                                  & 17401  & ···   \\
{[}4.5{]}                & 4.4491  &  & $1.276\pm0.015$  & $1.198\pm0.042$                                                     & $1.293\pm0.032$                                                      & $1.259\pm0.023$  & ···              &  & 3765  & 2901                                                                & 93                                                                   & 8561   & ···   \\
{[}5.8{]}                & 5.6597  &  & $1.305\pm0.015$  & $1.273\pm0.027$                                                     & $1.316\pm0.071$                                                      & $1.299\pm0.022$  & ···              &  & 2892  & 2379                                                                & 91                                                                   & 7490   & ···   \\
{[}8.0{]}                & 7.6738  &  & $1.352\pm0.015$  & $1.300\pm0.025$                                                     & $1.373\pm0.077$                                                      & $1.309\pm0.031$  & ···              &  & 1222  & 1834                                                                & 81                                                                   & 2966   & ···

\end{tabular}
\rule{1\linewidth}{1.4pt} 
\end{minipage}
}

\end{table}

\begin{table}[]
\centering
\caption{Multi-wavelengt Relative Extinction}
\label{tab:Relative Extinction}
\setlength{\tabcolsep}{11.5pt} 
\begin{tabular}{@{}cccccccc@{}}
\toprule
\midrule

Band                     & $\lambda_{\text{eff}, 0}$ &  & \multicolumn{5}{c}{$A_\lambda /A_\mathrm{v}$}                                                                                                                                           \\
\cline{4-8}
                         & ($\mu$m)                      &  & Coalsack               & $E_{G_\mathrm{BP},G_\mathrm{RP}}$$<1.25$ & $E_{G_\mathrm{BP},G_\mathrm{RP}}\geq1.25$ & Ref. 1              & Ref. 2              \\\midrule
$G_\mathrm{BP}$ & 0.5441                    &  & $1.040\pm0.001$ & ···                                                                 & ···                                                                  & $1.065\pm0.001$ & $1.057\pm0.001$ \\
$G_\mathrm{RP}$ & 0.7927                    &  & $0.605\pm0.001$ & ···                                                                 & ···                                                                  & $0.620\pm0.001$ & $0.616\pm0.001$ \\
$\mathit{B}$                        & 0.4487                    &  & $1.287\pm0.002$ & ···                                                                 & ···                                                                  & $1.304\pm0.002$ & $1.313\pm0.002$ \\
$\mathit{V}$                        & 0.5519                    &  & $1.000\pm0.001$ & $1.000\pm0.001$                                                     & $1.000\pm0.001$                                                      & $1.000\pm0.001$ & $1.000\pm0.001$ \\
$\mathit{J}$                        & 1.2373                    &  & $0.271\pm0.001$ & $0.273\pm0.001$                                                     & $0.254\pm0.002$                                                      & $0.294\pm0.001$ & $0.306\pm0.001$ \\
$\mathit{H}$                        & 1.6442                    &  & $0.144\pm0.001$ & $0.147\pm0.001$                                                     & $0.147\pm0.002$                                                      & $0.164\pm0.001$ & $0.171\pm0.001$ \\
$\mathit{K_S}$           & 2.1567                    &  & $0.099\pm0.001$ & $0.098\pm0.001$                                                     & $0.129\pm0.002$                                                      & $0.105\pm0.001$ & $0.111\pm0.001$ \\
$\mathit{u}$                        & 0.3517                    &  & $1.501\pm0.002$ & ···                                                                 & ···                                                                  & $1.604\pm0.002$ & $1.587\pm0.002$ \\
$\mathit{v}$                        & 0.3854                    &  & $1.497\pm0.001$ & ···                                                                 & ···                                                                  & $1.584\pm0.002$ & $1.602\pm0.002$ \\
$\mathit{g}$                        & 0.5148                    &  & $1.104\pm0.000$ & $1.110\pm0.001$                                                     & $1.061\pm0.001$                                                      & $1.130\pm0.001$ & $1.117\pm0.000$ \\
$\mathit{r}$                        & 0.6160                    &  & $0.871\pm0.001$ & $0.875\pm0.001$                                                     & $0.846\pm0.001$                                                      & $0.898\pm0.001$ & $0.885\pm0.001$ \\
$\mathit{i}$                        & 0.7767                    &  & $0.621\pm0.001$ & $0.621\pm0.001$                                                     & $0.603\pm0.001$                                                      & $0.622\pm0.001$ & $0.615\pm0.001$ \\
$\mathit{z}$                        & 0.9135                    &  & $0.473\pm0.001$ & $0.470\pm0.001$                                                     & $0.459\pm0.001$                                                      & $0.481\pm0.001$ & $0.474\pm0.001$ \\\midrule
                         &                           &  & \multicolumn{5}{c}{$A_\lambda / A_{K_\mathrm{s}}$}                                                                                                                    \\
\cline{4-8}
                  &                           &  &    Coalsack            & $ E_{J,K_s}<0.5$                                  & $E_{J,K_s}\geq0.5$                                  &      Ref. 1         &  Ref. 2         \\\midrule
$\mathit{J}$                       & 1.2373                    &  & $2.737\pm0.001$ & $2.795\pm0.001$                                                     & $2.839\pm0.002$                                                      & $2.800\pm0.001$ & $2.757\pm0.001$ \\
$\mathit{H}$                        & 1.6442                    &  & $1.687\pm0.001$ & $1.605\pm0.001$                                                     & $1.421\pm0.002$                                                      & $1.696\pm0.001$ & $1.495\pm0.001$ \\
$\mathit{K_S}$           & 2.1567                    &  & 1               & 1                                                                   & 1                                                                    & 1               & 1               \\
$\mathit{W1}$                       & 3.3397                    &  & $0.634\pm0.002$ & $0.745\pm0.002$                                                     & $0.778\pm0.003$                                                      & $0.638\pm0.002$ & $0.814\pm0.002$ \\
$\mathit{W2}$                        & 4.5811                    &  & $0.511\pm0.002$ & $0.691\pm0.002$                                                     & $0.699\pm0.003$                                                      & $0.555\pm0.002$ & $0.763\pm0.001$ \\
$\mathit{W3}$                        & 11.9151                   &  & $0.418\pm0.004$ & ···                                                                 & ···                                                                  & $0.493\pm0.004$ & $0.560\pm0.004$ \\
{[}3.6{]}                & 3.5205                    &  & $0.512\pm0.001$ & $0.649\pm0.001$                                                     & $0.785\pm0.002$                                                      & $0.498\pm0.001$ & ···             \\
{[}4.5{]}                & 4.4491                    &  & $0.520\pm0.001$ & $0.644\pm0.001$                                                     & $0.716\pm0.002$                                                      & $0.534\pm0.001$ & ···             \\
{[}5.8{]}                & 5.6597                    &  & $0.469\pm0.001$ & $0.509\pm0.001$                                                     & $0.694\pm0.002$                                                      & $0.461\pm0.001$ & ···             \\
{[}8.0{]}                & 7.6738                    &  & $0.389\pm0.001$ & $0.461\pm0.001$                                                     & $0.314\pm0.002$                                                      & $0.444\pm0.001$ & ···            
\end{tabular}
\rule{1\linewidth}{0.7pt} 
\end{table}

\end{CJK}
\end{document}